# Negative redispatch power for green hydrogen production: Game changer or lame duck? A German perspective

J. Brandt*,1, A. Bensmann*,1, R. Hanke-Rauschenbach*

*Leibniz Universität Hannover, Institute for Electric Power Systems, 30167 Hanover, Germany

[1]Corresponding authors (brandt@ifes.uni-hannover.de, astrid.bensmann@ifes.uni-hannover.de)

## Abstract

Following years of controversial discussions about the risks of market-based redispatch, the German transmission network operators finally installed regional redispatch markets by the end of 2024. Since water electrolysers are eligible market participants, the otherwise downwards redispatched renewable energy can be used for green hydrogen production in compliance with European law. To show how different price levels in regional redispatch markets affect green hydrogen production cost and thus the incentive for electrolyser market participation, we use historic redispatch time series and evaluate various power purchase scenarios. Our results show that low price levels can lead to notable production cost reductions, potentially counteracting uncertainties in redispatch power availability and thus incentivising system-beneficial electrolyser siting. In contrast, the possibility of high price levels can nullify an increase in the competitiveness of German and European green hydrogen through production cost reductions and discourage market participation.

## Keywords

Green hydrogen production, regional redispatch markets, system-beneficial electrolyser siting, relief regions, local price signals

## 1 Introduction

With the further deployment of renewable energy generation in Europe to reach emission reduction targets, its curtailment due to grid limitations is expected to increase heavily in the upcoming years and reach between 111 and 310 TWh in 2040 [1]. In Germany, downwards redispatched renewable energy has risen for years, adding up to 10 TWh in 2023 alone [2]. Despite the European Union obligations to install a market-based redispatch to make the curtailed renewable energy usable in a market environment [3], German authorities resisted moving away from the established cost-based redispatch for years [4]. The reasons given were mirrored by controversial discussions about feared strategic and congestion-enhancing market behaviour and potentially insufficient competition in regional redispatch markets [5]–[8]. In the end, the German parliament implemented European Union law in late 2023 and passed §13k EnWG "Nutzen statt Abregeln 2.0" [9], [10]. The passed regulation obliges transmission network operators to make otherwise downwards redispatched renewable energy accessible for electrical loads as hourly products in regional markets upstream of the day-ahead market [9]–[11]. Expected system benefits are reducing renewable energy curtailments and relief of grid congestions and congestion management costs [9]–[11]. A two-year test phase for the regional redispatch markets started in October 2024 [11].

Since electrolysers are eligible market participants [11], a new power purchase option for renewable (henceforth, green) hydrogen production opens up [12], [13]. Therefore, electrolyser participation could not only contribute to the markets' success and the associated system benefits but also decrease green hydrogen production cost, and thus accelerate the slowed ramp-up of European and German green hydrogen production [14]–[17]. However, it remains unclear whether, at all, and at which market price levels redispatch power integration would result in sufficient green hydrogen supply cost reductions to incentivise regional electrolyser siting and market participation.

Numerous studies investigated different factors influencing green hydrogen supply cost, whereby linear optimisation of the design and operation of electrolyser plants turned out to be a suitable methodological choice [13], [18]–[23]. Additionally noteworthy in the context of the present study are analyses on European [24], [25] and German [26]–[28] system level showing the system-beneficial effects of strategic electrolyser siting. One study additionally highlights that inefficient siting could even increase renewable energy curtailment [1]. Although transmission network operators would be obvious players for system optimal electrolyser siting and operation [29], German authorities have rejected requests in this regard for unbundling reasons [30]–[32]. Therefore, [1], [25], [27] point out the necessity to incentivise system optimal electrolyser siting and operation by industry players. According to [1], [33], [34], local price signals through price zone splitting or nodal pricing could be suitable instruments in this regard.

To answer the question whether and at which price levels the recently introduced German regional redispatch markets are a suitable instrument to incentivise system-beneficial electrolyser siting and market participation, this work consists of the following steps. First, we introduce a set of historic redispatch time series for all German regional market areas in 2022 and 2023 in section 2. Second, those time series provide input data to a linear optimisation model used to determine the minimal onsite hydrogen supply cost of a hydrogen production plant with different power purchase options presented in section 3. Subsequently, different power purchase scenarios including and excluding a power purchase option from regional redispatch markets are introduced in section 4. Then, the minimal achievable hydrogen supply costs in those different power purchase scenarios are compared to quantify possible cost reductions associated with redispatch power integration in sections 5.1, 5.2 and 5.4. The analyses were repeated for different redispatch market price levels and storage options to allow a comprehensive assessment of whether and at which price levels the regional redispatch markets provide sufficient incentive for electrolyser participation. In addition, a deep dive analysis in section 5.3 shows how forward-looking PPA and component sizing enables the hydrogen supply cost reductions from redispatch usage. Finally, whether and to what extend the findings are transferable to other European countries is addressed in section 6.

## 2 Redispatch time series

To be able to mathematically map the availability of otherwise downwards redispatched renewable power in all regional German redispatch markets, we recreated historic redispatch time series for the years 2022 and 2023. For this purpose, we combined bias-corrected reanalysis weather data with reference plant models of renewable energy sources (RES) and redispatch information from all relevant transmission and distribution system operators in northern Germany. Finally, the resulting hourly resolved redispatch power time series for over 1000 transformer stations were summed up within eight initially defined geographical market areas, referred to as relief regions (RR) in the following (German original term: Entlastungsregionen) and shown in Figure 1 (a). Supplementary Note 1 provides further details on the redispatch time series' creation and the validation process mentioned in the following paragraphs.

Figure 1 (b) and (c) depict the recreated dataset, whereby the grey dots are the single transformer station data, and the colored dots are the geographically combined data inside the RRs. Figure 1 (b) shows the number of hours in the time series with a redispatch measure and Figure 1 (c) the equivalent full load hours of the time series. Each above the respective energy amount sum of the time series. Although Figure 1 (b) shows that the summation inside the RRs leads to increased time steps with redispatch measures and higher energy amounts compared to most transformer station data, the resulting equivalent full load hours show no recognizable increase. It is important to note that the transformer station data maps the whole network areas of relevant system operators. This also includes locations that are not inside any RR, which are therefore excluded from the summation inside the RRs.

Comparing the recreated transformer station redispatch time series for validation purposes with publications from German authorities and transmission system operators shows that they cover more than 98 % of downwards redispatched renewable energy amounts on transmission level and more than 76 % on distribution level in both 2022 and 2023.

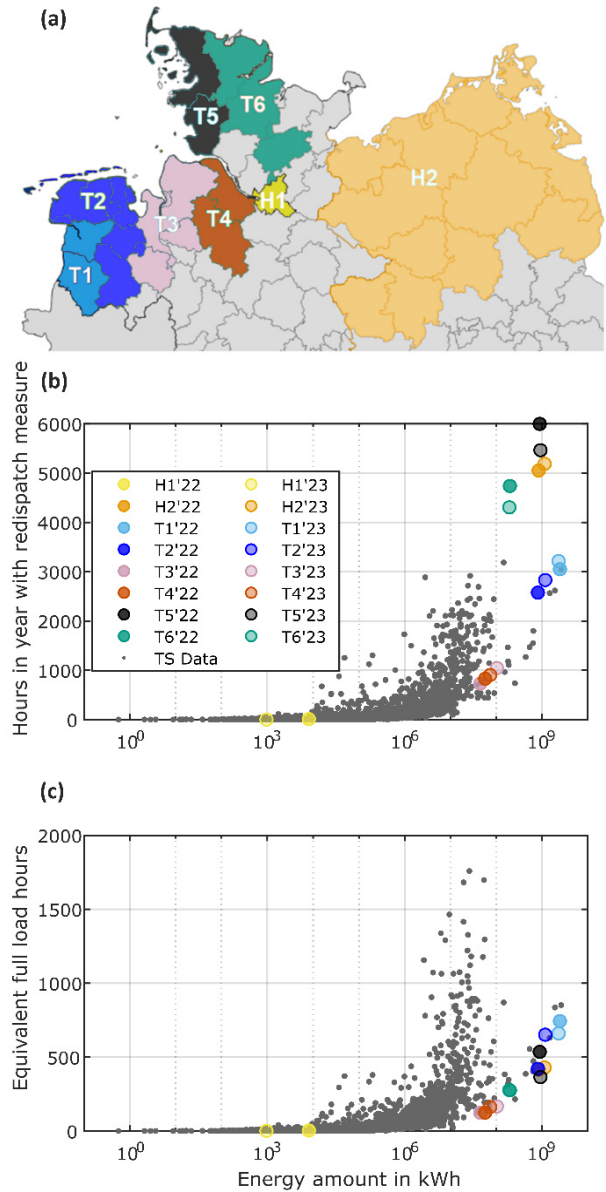

Figure 1: Redispatch time series. (a), Regional German redispatch market areas named as in [11] *but recolored. (b)-(c), Transformer station related (grey dots) and geographically inside the RRs combined (colored dots) redispatch time series and their key characteristics. (b), Number of hours in year with a redispatch measure above the respective energy amount sum of the time series. (c), Equivalent full load hours of redispatch time series above the respective energy amount sum of the time series*

## 3 Methodological approach

To evaluate the economic viability of different power purchase scenarios, including and excluding redispatch power, the incurring cost of the hydrogen production system in Figure 2 to meet a predefined hydrogen demand is mathematically minimised (equation (1)). By this means, we ensure to always compare cost-optimal plant sizing and dispatch for all analysed scenarios, hydrogen storage options and redispatch price variations.

$$\min \sum_{i=1}^{N}(C_{\text{TAC,CAPEX},i} + C_{\text{TAC,OPEX},i}) + \sum_{j=1}^{M} C_{\text{PP},j} \qquad (1)$$

The minimised cost is composed of the annualized installation cost $C_{\text{TAC,CAPEX},i}$ and the annual operation cost $C_{\text{TAC,OPEX},i}$ of all components $N$ of the hydrogen production plant and the power purchase cost $C_{\text{PP},j}$ of all power purchase options $M$. The onsite hydrogen supply costs (OHSC) are calculated by dividing the minimised annual cost by the annual sum of the predefined hydrogen demand (equation (2)). They serve as an index usable for comparing the economic viability of hydrogen production in the different power purchase scenarios.

$$\text{OHSC} = \frac{\sum_{i=1}^{N} C_{\text{TAC,CAPEX},i} + C_{\text{TAC,OPEX},i} + \sum_{j=1}^{M} C_{PP,j}}{\sum_{t=1}^{T} \dot{m}_{\text{demand},t} \cdot \Delta t} \qquad (2)$$

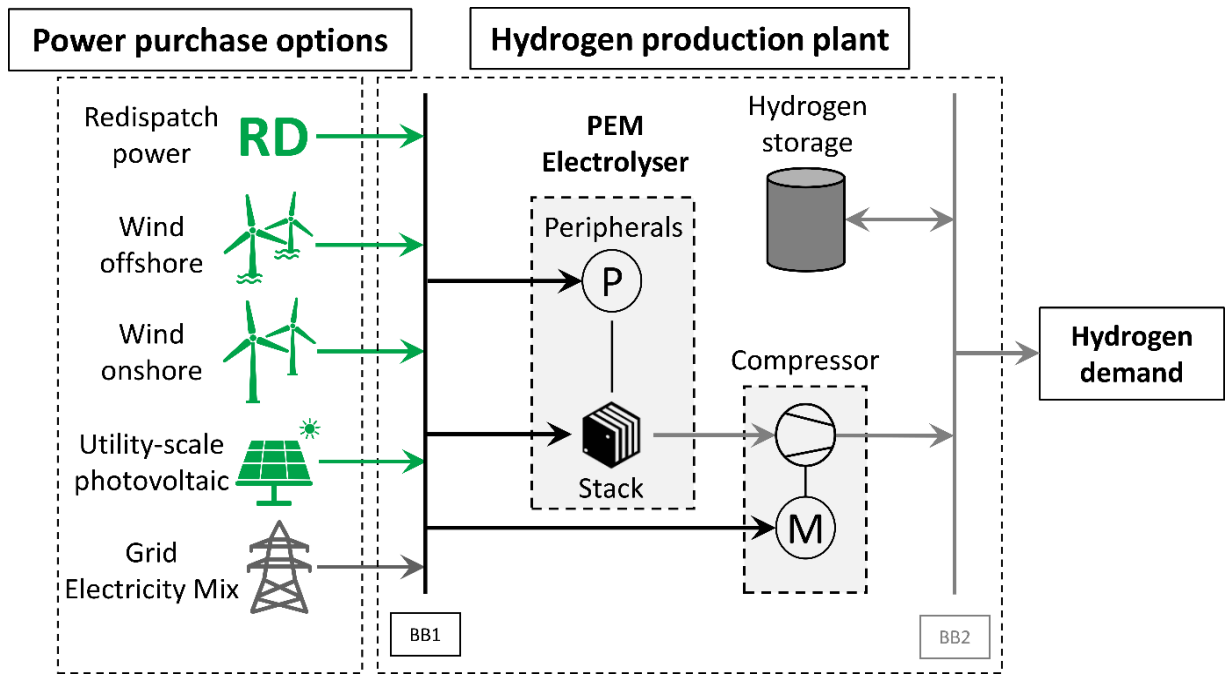

Figure 2: Set-up of the modelled hydrogen production system. The dashed box on the left shows all available power purchase options including the option to purchase redispatch power from a regional market. The dashed box on the right shows the hydrogen production plant consisting of a proton exchange membrane (PEM) electrolyser including peripherals (P) and stack, a piston compressor powered by an electric motor (M), and a hydrogen storage optionally as pressure gas tank, salt cavern storage bundle or free-of-charge storage option. Vertical lines: BB1, electricity bus bar; BB2, hydrogen bus bar.

The power purchase agreement (PPA) options wind offshore, wind onshore, and utility-scale photovoltaic are pay-as-produced power purchase options. Supplementary Note 2 and 3 provide further details on PPA modeling. For redispatch power and grid electricity mix purchase fixed prices are assumed, whereby the price level for electricity mix purchase is the average German electricity price for non-household consumers in the first half of 2024 [35]. The redispatch price levels are the subject of this study. To increase the comprehensibility of this study and to be able to make general findings, we simplified the quite complex pricing mechanism of the German redispatch markets. Since there are no electrical grid fees and taxes of a notable size for electrolyser operation in Germany, they are assumed to be zero [36], [37]. For electrolyser and compressor sizing, annualized installation cost incur. The analysed hydrogen storage options are, installing a pressure gas tank, the allocation of a salt cavern storage bundle, and a free-of-charge storage option. The free-of-charge option makes it possible to investigate limiting cases, which occur when there are no requirements regarding hydrogen delivery time from the customer or if a free-of-charge buffer option, like an oversized hydrogen pipeline network, would be available. The time characteristic of the predefined hydrogen demand is chosen to be flat, mirroring the demand of a large-scale industrial costumer from the steel or chemical sector. The variation of its annual sum is part of the following analyses.

Supplementary Note 2 lists all economic and technical parameters. For this study, all economic input parameters are set to present cost values derived from recent publications and converted into euros (2024 value, €$_{2024}$) using the Chemical Engineering Plant Cost Index (CEPCI) [38]. The formulated optimisation problem, including objective function and constraints, is linear, and the optimisation time frame is one year with an hourly resolution. Section 7.1 includes the complete mathematical formulation. By considering the sizing and dispatch of all hydrogen components in the cost minimisation, we determine the maximal achievable supply cost reductions from redispatch power integration. Different approaches, with fixed electrolyser project sizes, for example, could also be viable choices for similar investigations.

## 4  Power purchase scenarios

To investigate the economic viability of redispatch power usage for green hydrogen production, we derive five power purchase scenarios shown in Figure 3 from the setup in Figure 2. Scenario *PPA Ref.* exclusively includes the three PPA options for power sourcing. It serves as an economic reference for the scenarios *Redispatch (RD) only,* where redispatch power is the exclusive power sourcing option, and *RD + PPA,* where a mix of redispatch and PPA power is available. Comparing each with the reference scenario *PPA Ref.* allows us to evaluate the economic viability of **Green hydrogen production exclusively from redispatch power** in Chapter 5.1 and of **Combining redispatch and PPA power purchase** for green hydrogen production in Chapter 5.2, respectively. In addition, we present the scenarios *First Mover (FM)* and *RD + FM,* where additional power sourcing of grid electricity mix is available. Their comparison in Chapter 5.4 allows evaluation of how well redispatch power integration works for **First mover Projects** whose power sourcing complies with the transitional rules of the European Union regulatory framework [12], [13]. Chapter 7.1 provides the implemented constraints mapping the specifics of each scenario mathematically.

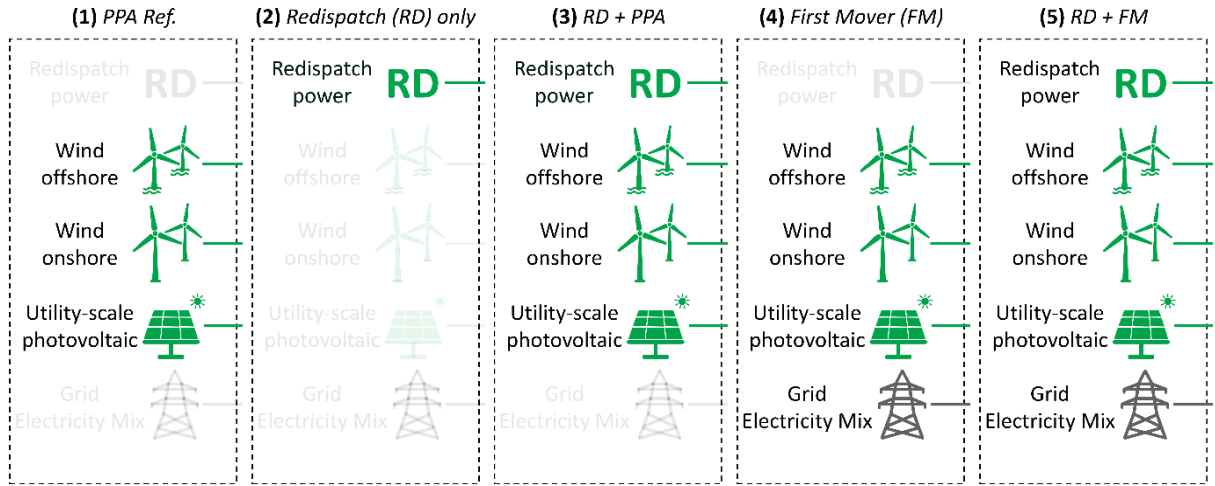

Figure 3: Analysed power purchase scenarios. (1), PPA Ref.. (2), Redispatch (RD) only. (3), RD + PPA. (4), First Mover (FM). (5), RD + FM.

## 5 Results and discussion

### 5.1 Green hydrogen production exclusively from redispatch power

To get a broad picture of the economic viability of green hydrogen production exclusively from redispatch power, the scenario *RD only* is computed for every RR and year and compared with the *PPA Ref.* scenario reference, given a variation of the annual sum of predefined hydrogen demand (see $\dot{m}_{\mathrm{demand},t}$ in equation (17)).

Figure 4 (a) shows the resulting OHSC and Figure 4 (b) the resulting nominal electrolyser power, resembling the project size. As an increase in hydrogen demand always leads to an increase in project size, although it is not always uniform, the terms are used analogously in the following. Both figures additionally depict grey star markers showing the results for a 300 MW project size. They serve as a guide to realistic project sizes, such as the *IPCEI* hydrogen projects currently being realised in some of the RRs [39]. The results in this first analysis are under the assumptions of a redispatch market price of 0 ct/kWh and the salt cavern storage bundle option. To increase the visibility of results, the RR H1 (cf. Figure 1) is excluded from this and all the following figures due to the lack of usability of the quasi-unavailable redispatch power in the RR for green hydrogen production (see Figure 1 (b) and (c) and all following).

The resulting OHSC in Figure 4 (a) are dependent on the annual sum of hydrogen demand, the associated project size, the specific RR in which the project is located, and the analysed year. Considering the 300 MW reference project size, for all RRs and years an OHSC above the reference in the *PPA Ref.* scenario results. It is between 0.1 €/kgH$_2$ in T1 in 2023 (dash-dotted light blue line in Figure 4 (a)) and a multiple of the reference in T4 in 2022 (solid red line in Figure 4 (a)) above the *PPA Ref.* OHSC. The right end of each graph shows the maximal supply cost resulting when all available redispatch power is needed to cover the predefined hydrogen demand. Comparing the resulting project sizes in Figure 4 (b) shows that exclusive redispatch power usage in all RRs necessitates an oversizing of nominal electrolyser power compared to the exclusive PPA power usage.

To get the most comprehensive impression of the effects of different redispatch market price levels on possible OHSC reductions, the computations were repeated for prices between 0 ct/kWh and 10 ct/kWh in 2 ct/kWh steps and for all available storage options. Figure 5 (a)-(i) show the OHSC reductions in each *RD only* scenario compared to the respective *PPA Ref.*

scenario for all storage options and redispatch prices. The fanning out of the graphs shows the redispatch price variation. The graphs show zero if the *RD only* scenario results in comparably higher OHSC. The rows of subfigures depict three exemplary RRs (T5, T1, T3), chosen due to their comprehensive representation of the time series characteristics shown in Figure 1 (b) and (c).

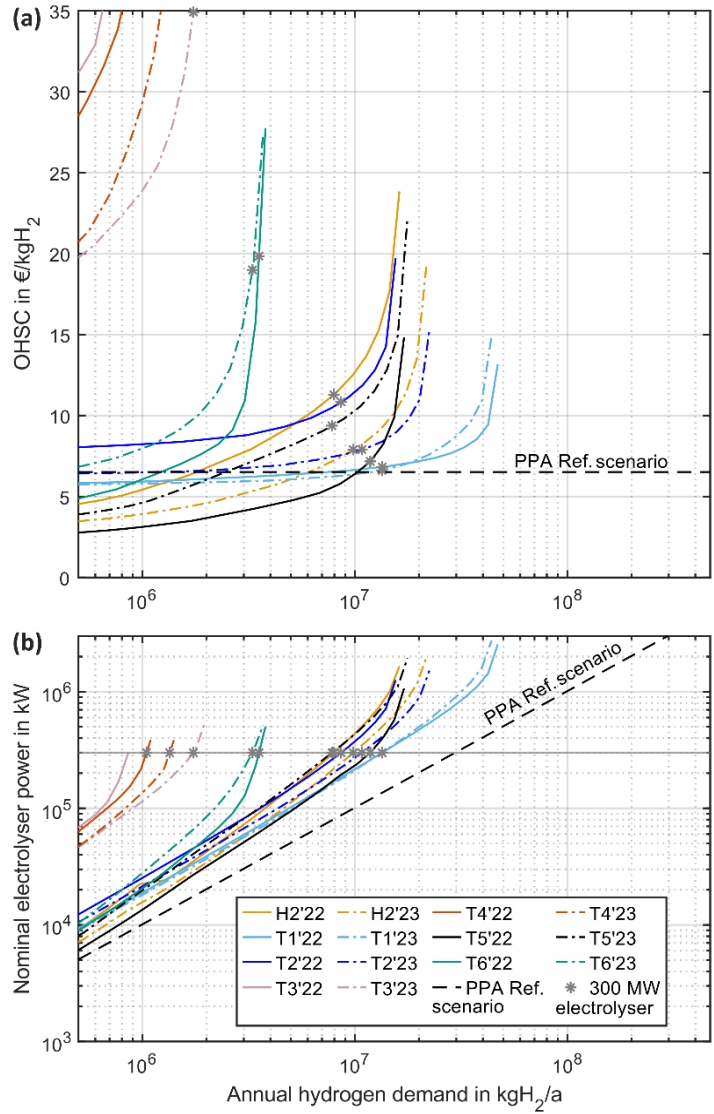

Figure 4: Optimisation results for green hydrogen production exclusively from redispatch power for all RRs (colour) and years (line style). (a), OHSC. (b), Nominal electrolyser power resembling the project size. The redispatch price is assumed to be 0 ct/kWh. The storage option is a salt cavern storage bundle. In addition, the grey star markers show the results for the reference project size of 300 MW. Furthermore, both graphs show the results of the PPA Ref. scenario as a reference (black dashed line).

Figure 5 (a)-(c) and (e)-(f) show that OHSC reductions through exclusive redispatch power usage compared to exclusive PPA power usage are generally possible. However, they occur almost exclusively for low redispatch prices, low storage costs and small project sizes. Considering the 300 MW project size, noticeable OHSC reductions are possible in T5 and T1 under the assumption of no storage cost (Figure 5 (c) and (f)). They range between 1.72 €/kgH$_2$ (solid light blue line at 0 ct/kWh in Figure 5 (f)) and 0.56 €/kgH$_2$ (dash-dotted light blue line at 2 ct/kWh in Figure 5 (f)) for redispatch prices of 0 ct/kWh and 2 ct/kWh. The inter-year comparison shows similar results in T1 (see Figure 5 (f)) but major differences in T5, where no OHSC reductions occur in 2023 (dash-dotted black lines in Figure 5 (c)) for any price level at the reference project size. For storage cost above zero and at the reference project size no OHSC reductions occur in any of the considered RRs (Figure 5 (a)-(b) and (d)-(e)). Figure 5 (g)-(i) show that in T3, no OHSC reductions are possible under any assumption. The same applies to T1 when assuming pressure tank storage cost (Figure 5 (d)). Supplementary Note 4 contains the results for all other regions.

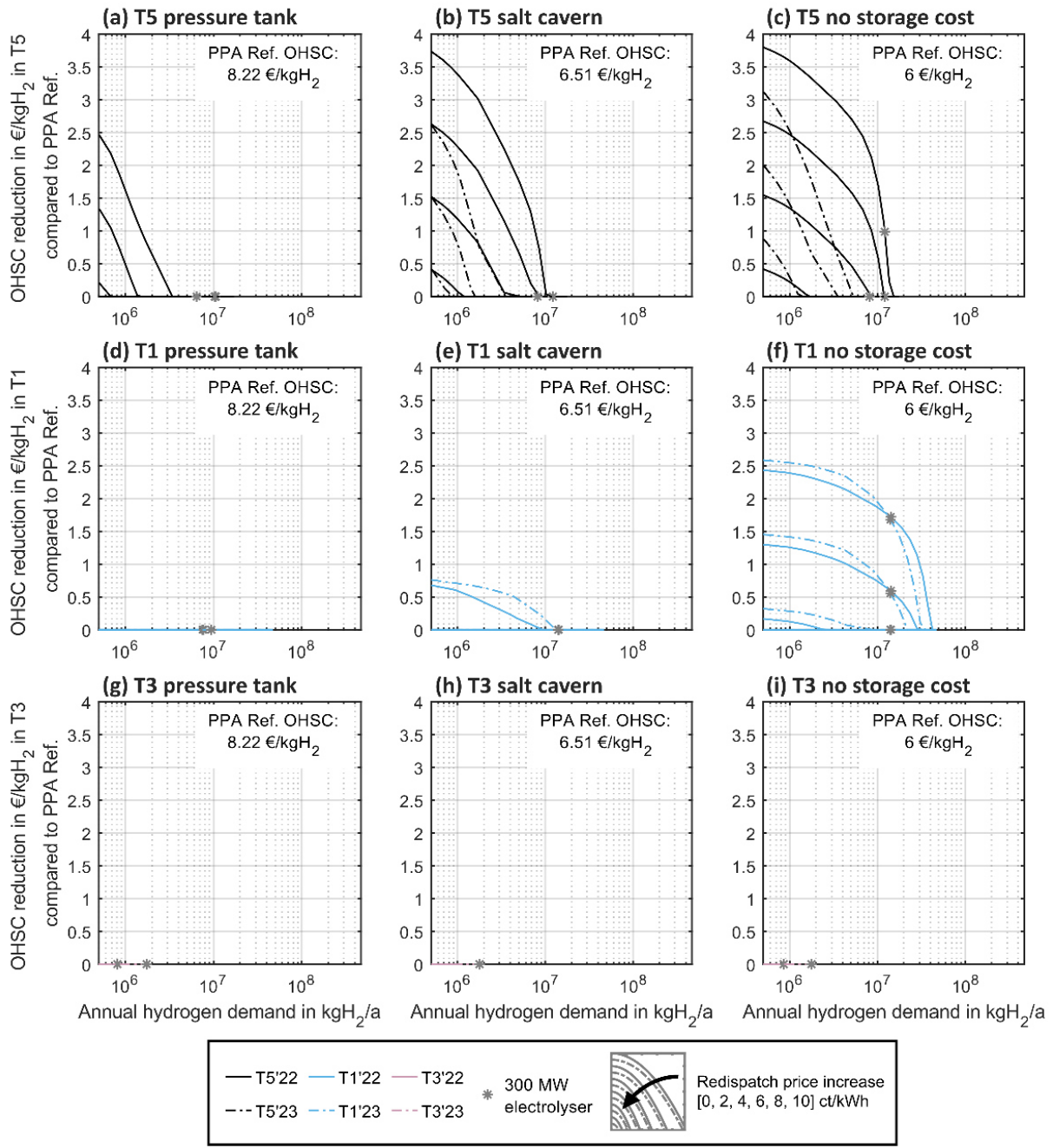

*Figure 5: Optimisation results for green hydrogen production exclusively from redispatch power for different redispatch price levels and storage options. (a)-(c), OHSC reductions from exclusive redispatch usage compared to exclusive PPA power sourcing in T5. (d)-(f), OHSC reductions from exclusive redispatch usage compared to exclusive PPA power sourcing in T1. (g)-(i), OHSC reductions from exclusive redispatch usage compared to exclusive PPA power sourcing in T3. In addition, the grey star markers show the results for the reference project size of 300 MW. Furthermore, all subfigures show the resulting OHSC of the respective PPA Ref. scenario as a reference.*

Taking the shown inter-year uncertainty regarding redispatch power availability and realistic electrolyser project sizes into account, the option to produce green hydrogen exclusively from redispatch power can be ruled out to be a cost-competitive option. This applies to all RRs, redispatch price levels, and all hydrogen storage options. The only exception is T1 in the limiting case if no storage cost incurs.

## 5.2 Combining redispatch and PPA power purchase

Since producing green hydrogen exclusively from redispatch power turned out to be a non-cost-competitive option, the economic viability of complementing PPA with redispatch power for green hydrogen production is assessed by analysing the combined power purchase scenario *PPA + RD*, again in comparison to *PPA Ref.*. Figure 6 shows the results. The assumptions made and the type of display are the same as in Figure 4.

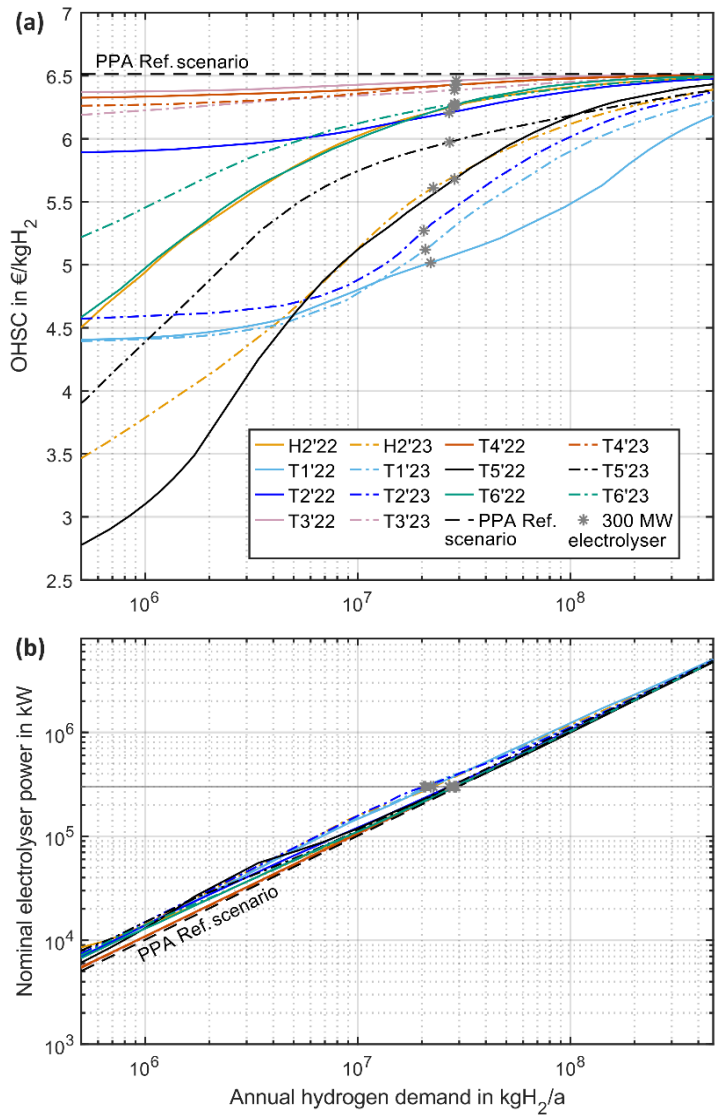

*Figure 6: Optimisation results for green hydrogen production from combined redispatch and PPA power for all RRs (colour) and years (line style). (a), OHSC. (b), Nominal electrolyser power resembling the project size. The redispatch price is assumed to be 0 ct/kWh. The storage option is a salt cavern storage bundle. In addition, the grey star markers show the results for the reference project size of 300 MW. Furthermore, both graphs show the results of the PPA Ref. scenario as a reference (black dashed line).*

Comparing the mix of PPA and redispatch power with the exclusive usage of PPA power shows lower OHSC in all RRs and years in Figure 6 (a). The magnitude of possible OHSC reductions depends on project size and the analysed year. Assuming the 300 MW electrolyser size, the OHSC range from 5.02 €/kgH$_2$ in T1 in 2022 (solid light blue line) to 6.46 €/kgH$_2$ in T3 in 2022 (solid pink line). Compared to the PPA reference at 6.51 €/kgH$_2$, this accounts for OHSC reductions between 23 % and 0.8 %, respectively. Additionally noticeable are the inter-year differences. For instance, for a project size of 300 MW in T2 in 2023 (dash-dotted dark blue line), 1.24 €/kgH$_2$ of OHSC reduction results, compared to just 0.3 €/kgH$_2$ in 2022 (solid dark blue line).

Similar to section 5.1, we repeated the computations for different price levels and storage options. Figure 7 depicts the equivalent to Figure 5 with OHSC reductions in the *PPA + RD* scenario compared to the *PPA Ref.* scenario.

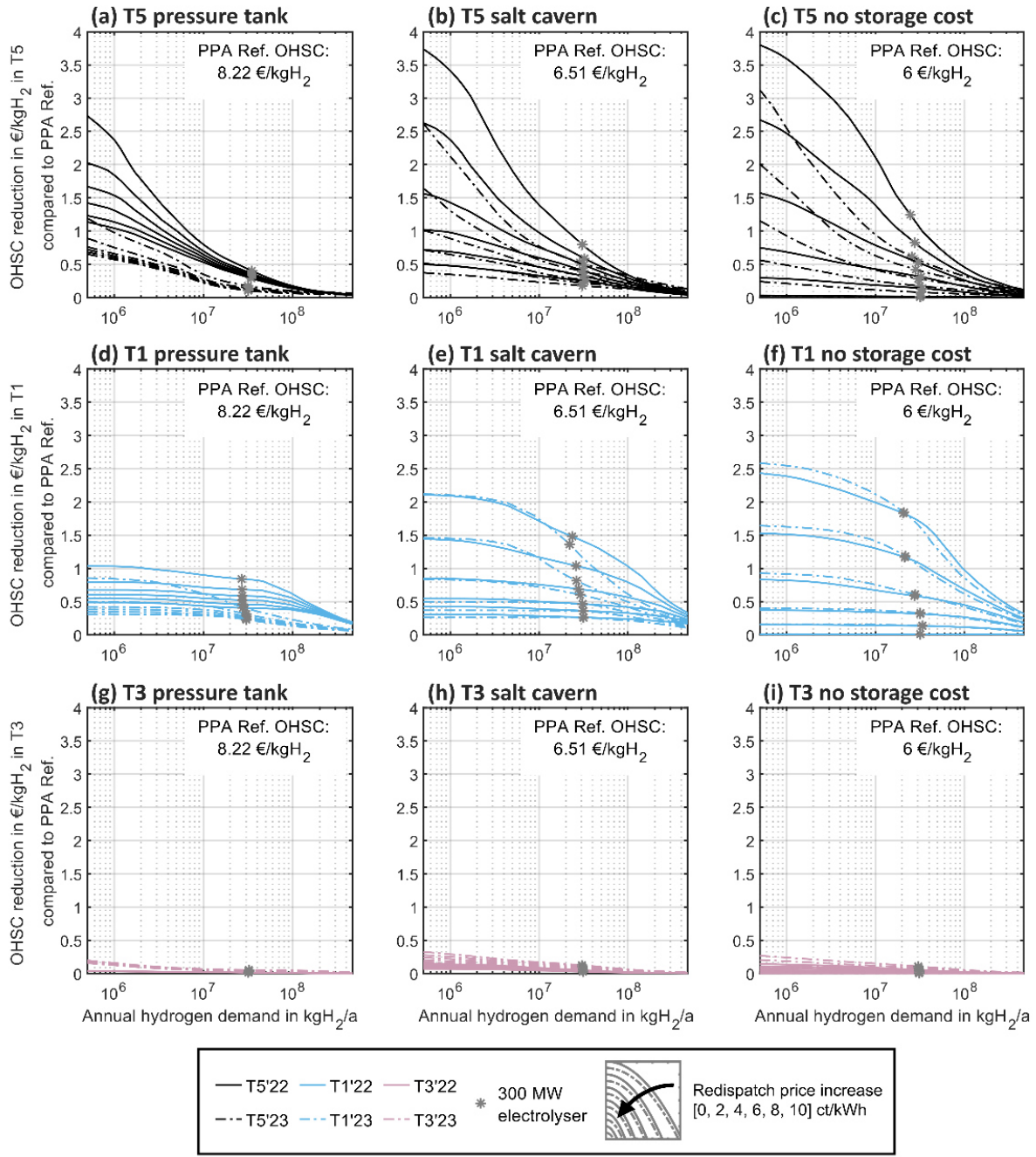

Figure 7: Optimisation results for green hydrogen production from combined redispatch and PPA power for different redispatch price levels and storage options. (a)-(c), OHSC reductions from combined PPA and redispatch power usage compared to exclusive PPA power sourcing in T5. (d)-(f), OHSC reductions from combined PPA and redispatch power usage compared to exclusive PPA power sourcing in T1. (g)-(i), OHSC reductions from combined PPA and redispatch power usage compared to exclusive PPA power sourcing in T3. In addition, the grey star markers

*show the results for the reference project size of 300 MW. Furthermore, all subfigures show the resulting OHSC of the respective PPA Ref. scenario as a reference.*

Figure 7 (a) – (f) show a dependency of OHSC reductions on both storage cost and redispatch prices. In both RRs T5 and T1, combining low storage costs and redispatch prices leads to the highest possible OHSC reductions. They are 1.24 €/kgH$_2$ in T5 in 2022 in Figure 7 (c) (solid black line) and 1.84 €/kgH$_2$ in T1 in 2022 in Figure 7 (f) (dash-dotted light blue line) at a 300 MW project size and a 0 ct/kWh redispatch price. In both RRs, increasing redispatch prices up to 10 ct/kWh leads to the complete elimination of possible OHSC reductions, contrasting the results of the pressure tank and salt cavern storage options (Figure 7 (a)-(b) and (d)-(e)). Here, the maximum achievable OHSC reduction at 0 ct/kWh redispatch price occurs in T1 in 2022 in Figure 7 (d) and (e) (solid light blue line) with just 0.84 €/kgH$_2$ and 1.48 €/kgH$_2$ respectively at 300 MW project size. However, the decrease in OHSC reductions with increasing redispatch prices never results in complete elimination. The magnitude of decrease declines with increasing storage cost and ranges between 82 % (1.48 to 0.26 €/kgH$_2$) in T1 in 2022 in Figure 7 (e) (solid light blue line) for salt cavern storage and 28 % (0.4 to 0.29 €/kgH$_2$) in T5 in 2022 in Figure 7 (a) (solid black line) for pressure tank storage at a 300 MW project size. Inter-year differences are noticeable in both RRs and for all hydrogen storage options. They are particularly striking in T5 in Figure 7 (c), where the maximum OHSC reduction at 0 ct/kWh redispatch price and 300 MW project size decreases by 50 % from 1.24 €/kgH$_2$ in 2022 (solid black line) to 0.62 €/kgH$_2$ in 2023 (dash-dotted black line). Contrastingly, T1 in Figure 7 (f) shows high inter-year similarities for all redispatch price levels (compare solid and dash-dotted light blue lines). In T3 only marginal OHSC reductions are achivable for all hydrogen storage options, all redispatch price levels and, in both years as depicted in Figure 7 (g)-(i). Especially at and above the 300 MW reference project, they are close to zero. Supplementary Note 5 contains the results for all other regions.

In summary, complementing PPA power purchase with redispatch power can be a valid option to increase the competitiveness of green hydrogen production. The magnitude of possible supply cost decrease depends on the redispatch price level, incurring hydrogen storage cost, the electrolyser project size, and inter-year differences in redispatch power availability in the different RRs. Hereby, the negative effect of increased redispatch prices is particularly noticeable if low cost storage options are available. Therefore, low redispatch price levels could provide an economic incentive to encourage system-beneficial electrolyser siting and market participation. On the other hand, the risk of high price levels in an unregulated market environment combined with the inter-year uncertainty about redispatch power availability could deter from market participation.

## 5.3 Details on cost reduction composition, Example: RR T5

To comprehend how the OHSC reductions come about, Figure 8 (a)-(i) show result figures for RR T5 in 2022 for different redispatch price levels and a salt cavern as hydrogen storage option. The black arrows indicate the direction in which redispatch price levels are increasing. Figure 8 (a) shows the OHSC reductions in T5 in 2022, as in Figure 7 (b). Figure 8 (b)-(c) show the share of power purchase from PPAs and the redispatch market, respectively. Figure 8 (d)-(f) show how the share of used redispatch energy in (d) contributes to the utilisation of the electrolyser in (e) and leads to pro rata OHSC reductions in (f). Figure 8 (g)-(i) show the pro rata OHSC reductions attributable to PPA, electrolyser and, storage sizing compared to the *PPA Ref.* scenario. Chapter 7.2 includes mathematical details on the calculation of the result figures.

Comparing Figure 8 (b) and (c) shows that low redispatch price levels and small project sizes lead to an increased share of redispatch power purchase. Despite an almost even decrease in OHSC reductions when increasing the redispatch price level from 0 ct/kWh to 4 ct/kWh in Figure 8 (a), the redispatch power share in Figure 8 (c) between 0 ct/kWh and 2 ct/kWh decreases disproportionately. It drops from 37 % to 13 % at the 300 MW project size. The share of used redispatch energy in Figure 8 (d) confirms this observation, with a decrease of 64 % from 69 % to 25 % for redispatch price levels above 0 ct/kWh. This affects the redispatch-associated electrolyser utilisation, which shows a reduction from 1946 to 708 full load hours for redispatch price levels above 0 ct/kWh (Figure 8 (e)).

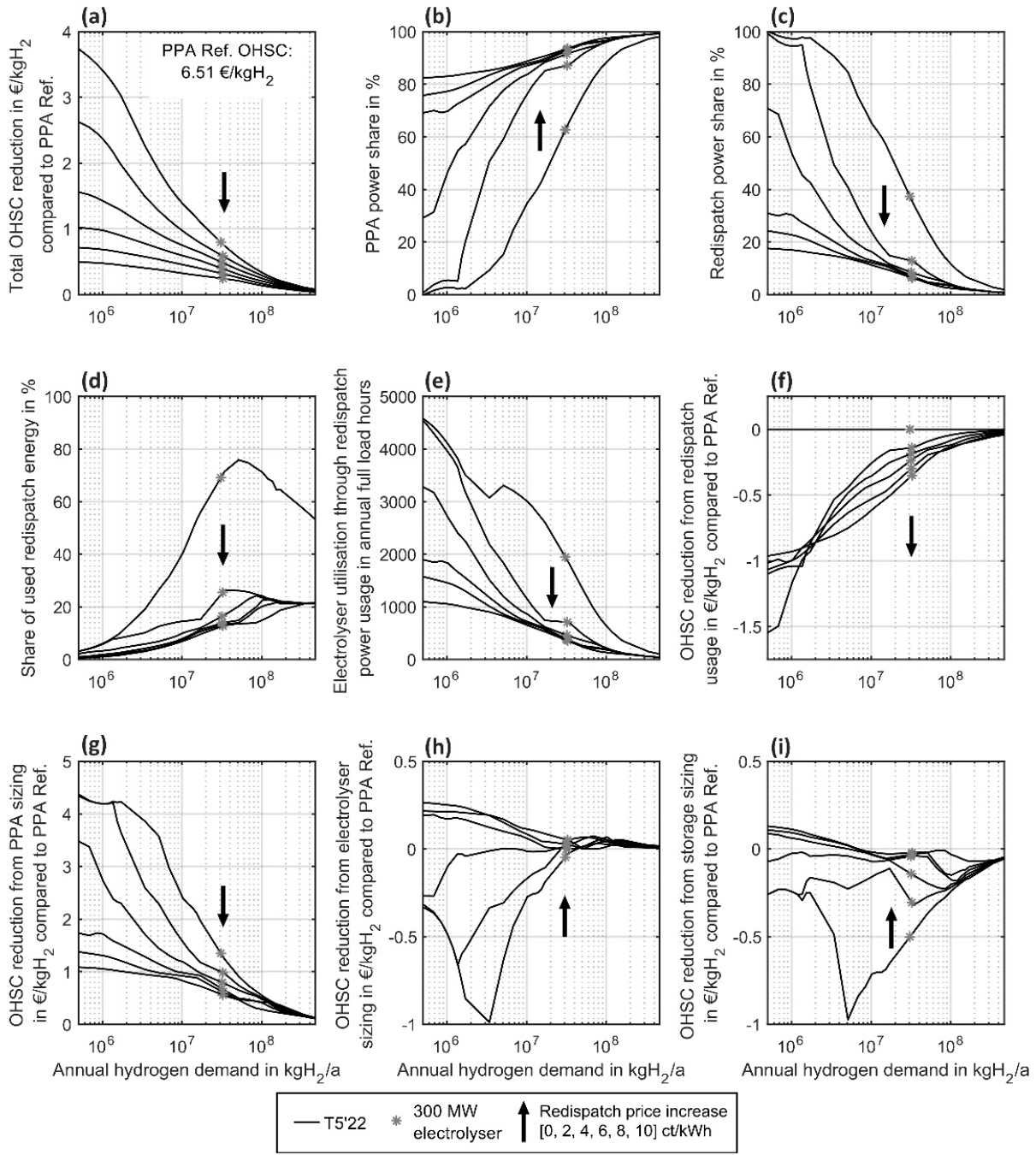

Figure 8: Deep dive into optimisation results with focus on composition of OHSC reductions by using result figures exemplified by T5 in 2022. (a), OHSC reductions from combined PPA and redispatch power usage compared to exclusive PPA power sourcing. (b)-(c), Share of power sourcing from PPAs and the redispatch market. (d), Share of used redispatch energy in available redispatch energy. (e), Electrolyser utilisation associated with redispatch power usage. (f), Pro rata OHSC reductions associated with redispatch power usage. (g), Pro rata OHSC reductions associated with PPA sizing. (h), Pro rata OHSC reductions associated with electrolyser sizing. (i), Pro rata OHSC reductions associated with hydrogen storage sizing. The black arrows indicate the direction of increasing redispatch

*price levels. The pressure tank is chosen as hydrogen storage option. In addition, the grey star markers show the results for the reference project size of 300 MW.*

The pro rata OHSC from redispatch power purchase above 0 ct/kWh in Figure 8 (f) range between 0.14 €/kgH$_2$ and 0.35 €/kgH$_2$. Comparing Figure 8 (g), (h), and (i) and considering (f) shows that the main contributor to OHSC reductions are PPA sizing adjustments compared to the *PPA Ref.* scenario. This applies to all redispatch price levels and project sizes. To achieve cost optimality, the PPA downsizing is combined with electrolyser (Figure 8 (h)) and storage (Figure 8 (i)) upsizing at project sizes below 300 MW and redispatch prices between 0 and 4 ct/kWh. Above 300 MW, just storage upsizing occurs at low price levels. For price levels above 4 ct/kWh, slight electrolyser (Figure 8 (h)) downsizing and storage (Figure 8 (i)) upsizing occurs, but sizing adjustments are less influential in general and become increasingly negligible above the reference project size.

The described effects are visible in all RRs, years and for all storage options. Even though the magnitude of effects differs from case to case, the main findings are generally transferable.

In conclusion, it can be stated that a redispatch price level of 0 ct/kWh incentivises an above average redispatch power usage. Furthermore, a forward-looking sizing of PPAs, electrolyser, and hydrogen storage enables OHSC reductions from redispatch power integration, whereby PPA sizing is the main driver. As PPA contract options are getting shorter, restructuring PPA power sourcing could allow maintaining OHSC reductions if major shifts in redispatch power availability occur.

## 5.4 First mover projects

Since the European Union regulatory framework allows limited usage of grid electricity mix for first mover green hydrogen projects until 2030, we repeated the preceding analyses for the *First Mover (FM)* and *FM + RD* scenarios. Although the additional power purchase option slightly changes the magnitude of possible OHSC reductions in single cases, the main findings are similar to those previously made and confirm the transferability of the conclusions drawn to first mover projects. Supplementary Note 6 includes the detailed results.

## 6 Conclusion

At the end of 2023, the German parliament converted European into national law and initiated the introduction of regional markets that make previously downwards redispatched renewable energy usable. Since the markets' test phase just started, it remains unclear how different price levels could affect green hydrogen supply cost and, thus, the incentive for regional siting and market participation of electrolysers.

In this paper, we used historic redispatch data to show that green hydrogen production exclusively from redispatch power is no cost-competitive option at any redispatch price level, for any relevant electrolyser project size in any regional redispatch market in Germany. Using redispatch power to supplement power purchase via PPAs, on the other hand, turns out to be a promising option that could decrease green hydrogen supply cost compared to exclusive PPA power usage. However, its economic viability relies on multiple factors. The magnitude of possible supply cost reductions depends on redispatch prices, the considered electrolyser project size, and incurring hydrogen storage cost and varies between regional markets.

Our results show that for an exemplary 300 MW electrolyser project size, the maximum achievable supply cost reductions for a redispatch price of 0 ct/kWh range from 0.9 €/kgH$_2$ for high hydrogen storage cost to 1.55 €/kgH$_2$ for low storage cost and 1.96 €/kgH$_2$ if no storage cost incurs. In most analysed scenarios, a forward-thinking downsizing of PPAs is the main driver for these supply cost reductions. Furthermore, a redispatch price of 0 ct/kWh, in particular, incentivises an above average redispatch power usage for green hydrogen production. Increasing the redispatch price above 0 ct/kWh not only results in an above average decrease in used redispatch power but also in a decrease of possible supply cost reductions of between 40 % and 100 % if the redispatch price reaches 10 ct/kWh. However, in some regional markets, considerable supply cost reductions cannot be achieved at any price level due to the geographical differences regarding available redispatch power. Additional analyses, including transitional rules for the power purchase of electrolysers in first mover projects, show similar results under the assumption of current German non-household consumer electricity prices.

Combined with the inter-year uncertainty about redispatch power availability in regional markets, the risk of increased price levels, possibly resulting from market competition, can decisively decrease hydrogen supply cost reductions from redispatch power usage. The resulting reduced economic incentive could outweigh the financial risks associated with system-beneficial electrolyser siting and adapted PPA sizing, thus detering from redispatch market participation. Guaranteed low redispatch market prices, on the other hand, could counteract the uncertainty of redispatch power availability and provide a compelling financial incentive for electrolyser siting and participation. This could finally contribute to the success of regional redispatch markets, leading to high usage rates of previously downwards redispatched renewable energy and associated system benefits. Simultaneously, decreased supply cost could increase the competitiveness of German and European green hydrogen and thus accelerate the slowed ramp-up of domestic green hydrogen production. Finally, the drawn conclusions raise the question of whether regional redispatch markets should be price- and access-regulated to ensure their success and anticipated system benefits.

We explicitly emphasize that the conclusions drawn in this study are in the context of the assumptions made. Critical aspects here are the chosen perspective of one single electrolyser operator and the simplification of market and pricing mechanisms, which allowed us to make general findings. In addition, the optimisation method leading to the maximum possible supply cost reductions should be mentioned due to the risk of overestimating realistically achievable supply cost reductions. Further aspects whose evaluation in subsequent studies could provide additional insights are the transferability of the findings obtained to the future and other European countries, especially due to the dependence of redispatch power availability on renewables and electricity grid expansion. However, the chosen methodological approach to map the access to a regional redispatch market as an additional fluctuating power source for electrolyser operation gives the drawn conclusions a general validity. Especially with the further increase of renewable energy curtailment and a potential introduction of similar markets in other European countries, the findings on how different price levels in regional redispatch markets incentivise system-beneficial electrolyser siting should provide transferable insights.

Due to the complexity of the subject and the mentioned limitations, this study alone cannot claim to provide an all-encompassing clarity on whether and how regional redispatch markets should be price- and access-regulated. Nevertheless, its findings should contribute to a better understanding of how different price levels in redispatch markets incentivise electrolyser siting and participation and thus could increase the competitiveness of German and European green hydrogen.

# 7 Appendix

## 7.1 Mathematical description of optimisation problem
To optimise design and operation of the chosen system setup in each scenario, the total annual cost of the system $C_{\text{tot}}$ is minimised (compare with equation (1)). It consists of the sum of annualized capital cost $C_{TAC,CAPEX,i}$ and total annual operation cost $C_{TAC,OPEX,i}$ for each component $i$ in the hydrogen production plant and the power purchase cost $C_{PP,j}$ for each power purchase option $j$ in Figure 2.

$$C_{tot} = \sum_{i=1}^{N}(C_{TAC,CAPEX,i} + C_{TAC,OPEX,i}) + \sum_{j=1}^{M} C_{PP,j} \qquad (3)$$

The annualized capital cost of each component in the hydrogen production plant is calculated by multiplying the nominal power $P_{nom,i}$ by the specific capital cost $c_{CAPEX,i}$ and the annuity payment factor $A_i$. Equation (5) shows the equivalent for hydrogen storage components.

$$C_{TAC,CAPEX,i} = P_{nom,i} \cdot c_{CAPEX,i} \cdot A_i \qquad (4)$$

$$C_{TAC,CAPEX,Sto} = m_{nom,Sto} \cdot c_{CAPEX,Sto} \cdot A_{Sto} \qquad (5)$$

The annuity payment factor is calculated as follows, where $r_{in,i}$ is the interest rate and $t_{dep,i}$ the depreciation time of the respective component. Hereby, the interest rate displays the nominal weighted average cost of capital.

$$A_i = \frac{r_{in,i} \cdot (1 + r_{in,i})^{t_{dep,i}}}{(1 + r_{in,i})^{t_{dep,i}} - 1} \qquad (6)$$

The annual operational cost of each component in the hydrogen production plant is calculated by multiplying the nominal power $P_{nom,i}$ by the specific capital cost $c_{CAPEX,i}$ and a fixed operational cost factor $f_{OPEXfix,i}$. To this fixed operational cost part a variable part is added by multiplying the annual sum of purchased power $P_{i,t}$ by the time step length and a variable operational cost factor $f_{OPEXvar,i}$, as shown in equation (7). Equation (8) shows the equivalent for hydrogen storage components.

$$C_{TAC,OPEX,i} = P_{nom,i} \cdot c_{CAPEX,i} \cdot f_{OPEXfix,i} + \sum_{t=1}^{T} P_{i,t} \cdot \Delta t \cdot f_{OPEXvar,i} \qquad (7)$$

$$C_{TAC,OPEX,Sto} = m_{nom,Sto} \cdot c_{CAPEX,Sto} \cdot f_{OPEXfix,Sto} + \sum_{t=1}^{T} \dot{m}_{StoIn,t} \cdot \Delta t \cdot f_{OPEXvar,Sto} \qquad (8)$$

The total cost for power purchase for power purchase option $j$ is calculated by multiplying the sum of purchased power $P_{j,t}$ of power purchase option $j$ for all time steps $t$ in the optimisation time frame $\{1,2,3,\ldots,T\}$ by the length of a time step $\Delta t$ and the specific power purchase price $p_j$.

$$C_{PP,j} = \sum_{t=1}^{T} P_{j,t} \cdot \Delta t \cdot p_j \qquad (9)$$

In case of PPAs, $P_{j,t}$ is replaced by the multiplication of the nominal power of the purchased PPA option $P_{nom,j}$ and the capacity factor $f_{cap,j,t}$ of the PPA option at time step $t$.

$$C_{PP,j} = \sum_{t=1}^{T} P_{nom,j} \cdot f_{cap,j,t} \cdot \Delta t \cdot p_j \qquad (10)$$

The following equations are the equality constraints that define the technical operation of the optimized hydrogen production system in Figure 2. All optimisation variables and parameters and their descriptions are listed in Table 1 and Table 2.

$$P_{RD,t} + P_{WindOff,t} + P_{WindOn,t} + P_{PV,t} + P_{Grid,t} = P_{Ely,t} + P_{Comp,t} \quad \forall t\epsilon\{1,2,3,\ldots,T\} \quad (11)$$

$$P_{Ely,t} = P_{Peri,t} + P_{Stack,t} \quad \forall t\epsilon\{1,2,3,\ldots,T\} \quad (12)$$

$$P_{ElySys,t} = P_{Ely,t} + P_{Comp,t} \quad \forall t\epsilon\{1,2,3,\ldots,T\} \quad (13)$$

$$P_{Peri,t} = \dot{m}_{Ely,t} \cdot \varepsilon_{Peri,t} \quad \forall t\epsilon\{1,2,3,\ldots,T\} \quad (14)$$

$$P_{Comp,t} = \dot{m}_{Ely,t} \cdot \varepsilon_{Comp,t} \quad \forall t\epsilon\{1,2,3,\ldots,T\} \quad (15)$$

$$\dot{m}_{H_2O,t} = \dot{m}_{Ely,t} \cdot \varepsilon_{H_2O,t} \quad \forall t\epsilon\{1,2,3,\ldots,T\} \quad (16)$$

$$\dot{m}_{Ely,t} \cdot (1 - f_{loss,Comp}) = \dot{m}_{StoIn,t} - \dot{m}_{StoOut,t} + \dot{m}_{demand,t} \quad \forall t\epsilon\{1,2,3,\ldots,T\} \quad (17)$$

$$m_{Sto,t=1} = m_{Sto,t=T} + (\dot{m}_{StoIn,t=1} - \dot{m}_{StoOut,t=1}) \cdot \Delta t \quad \forall t\epsilon\{1,2,3,\ldots,T\} \quad (18)$$

$$m_{Sto,t} = m_{Sto,t-1} + (\dot{m}_{StoIn,t} - \dot{m}_{StoOut,t}) \cdot \Delta t \quad \forall t\epsilon\{2,3,\ldots,T\} \quad (19)$$

The following equations are the inequality constraints that connect the operation and design of all relevant components and power purchase options.

$$P_{WindOff,t} \leq P_{nom,WindOff} \cdot f_{cap,WindOff,t} \quad \forall t\epsilon\{1,2,3,\ldots,T\} \quad (20)$$

$$P_{WindOn,t} \leq P_{nom,WindOn} \cdot f_{cap,WindOn,t} \quad \forall t\epsilon\{1,2,3,\ldots,T\} \quad (21)$$

$$P_{PV,t} \leq P_{nom,PV} \cdot f_{cap,PV,t} \quad \forall t\epsilon\{1,2,3,\ldots,T\} \quad (22)$$

$$P_{RD,t} \leq P_{RDava,t} \quad \forall t\epsilon\{1,2,3,\ldots,T\} \quad (23)$$

$$P_{Ely,t} \leq P_{nom,Ely} \quad \forall t\epsilon\{1,2,3,\ldots,T\} \quad (24)$$

$$P_{Stack,t} \leq P_{nom,Stack} \quad \forall t\epsilon\{1,2,3,\ldots,T\} \quad (25)$$

$$P_{Peri,t} \leq P_{nom,Peri} \quad \forall t\epsilon\{1,2,3,\ldots,T\} \quad (26)$$

$$P_{Comp,t} \leq P_{nom,Comp} \quad \forall t\epsilon\{1,2,3,\ldots,T\} \quad (27)$$

$$m_{Sto,t} \leq m_{nom,Sto} \quad \forall t\epsilon\{1,2,3,\ldots,T\} \quad (28)$$

To map the decrease in specific energy consumption of the electrolyser stack at operating points below the nominal load while keeping the mathematical formulation of the optimisation problem linear, the linearization method as in [13] was used. The respective constraints to construct a convex search space to keep the optimisation problem linear are the following.

$$\dot{m}_{Ely,t} \leq a_{lin,k} \cdot P_{Stack,t} + b_{lin,k} \quad \begin{array}{c} \forall t\epsilon\{1,2,3,\ldots,T\} \text{ and} \\ \forall k\epsilon\{1,2,3,\ldots,K-1\} \end{array} \quad (29)$$

with

$$a_{\text{lin},k} = \frac{k+1}{\varepsilon_{\text{Stack},k+1}} - \frac{k}{\varepsilon_{\text{Stack},k}} \tag{30}$$

$$b_{\text{lin},k} = \frac{\frac{k}{K-1} \cdot P_{\text{nom,Stack}}}{\varepsilon_{\text{Stack},k}} - a_{\text{lin},k} \cdot \frac{k}{K-1} \cdot P_{\text{nom,Stack}} \tag{31}$$

Hereby, $b_{\text{lin},k}$ is the y-axis intersect of the respective linear constraint $k$. The number of total linearization steps $K-1$ defines the number of additional inequality constraints needed per time step $t$. The number of linearization steps chosen in this study is 20. The inequality constraint in Equation (32) defines the lower bound of the constructed search space and is used to reduce the optimisation time by minimising the search space.

$$P_{\text{Stack},t} \leq \dot{m}_{\text{Ely},t} \cdot \varepsilon_{\text{nom,Stack}} \qquad \forall t \in \{1,2,3,\ldots,T\} \tag{32}$$

Equations (33) and (34) show the additional constraints for the First Mover scenarios *FM* and *FM + RD* respectively. Hereby the sum of electrolyser system power (*FM + RD*: minus purchased redispatch power) in month $l$ is not allowed to exceed the sum of contracted PPA power in the respective month.

$$\sum_{t=l\cdot 730+1}^{(l+1)\cdot 730} P_{\text{ElySys},t} \leq \sum_{t=l\cdot 730+1}^{(l+1)\cdot 730} (P_{\text{nom,WindOff}} \cdot f_{\text{capWindOff},t} + P_{\text{nom,WindOn}} \cdot f_{\text{capWindOn},t} + P_{\text{nom,PV}} \cdot f_{\text{capPV},t}) \qquad \forall l \in \{0,1,2,\ldots,11\} \tag{33}$$

$$\sum_{t=l\cdot 730+1}^{(l+1)\cdot 730} (P_{\text{ElySys},t} - P_{RD,t}) \leq \sum_{t=l\cdot 730+1}^{(l+1)\cdot 730} (P_{\text{nom,WindOff}} \cdot f_{\text{capWindOff},t} + P_{\text{nom,WindOn}} \cdot f_{\text{capWindOn},t} + P_{\text{nom,PV}} \cdot f_{\text{capPV},t}) \qquad \forall l \in \{0,1,2,\ldots,11\} \tag{34}$$

*Table 1: Optimisation variables*

| Variable | Description |
| --- | --- |
| $m_{\text{nom,Sto}}$ | Nominal capacity of hydrogen storage |
| $m_{\text{Sto},t}$ | Stored hydrogen mass at time step $t$ |
| $\dot{m}_{\text{Ely},t}$ | Hydrogen output mass flow of electrolyser plant at time step $t$ |
| $\dot{m}_{\text{StoIn},t}$ | Hydrogen mass flow into hydrogen storage at time step $t$ |
| $\dot{m}_{\text{StoOut},t}$ | Hydrogen mass flow out of hydrogen storage at time step $t$ |
| $P_{\text{nom,Comp}}$ | Nominal power of compressor |
| $P_{\text{Comp},t}$ | Compressor power consumption at time step $t$ |
| $P_{\text{nom,Ely}}$ | Nominal power of electrolyser |
| $P_{\text{Ely},t}$ | Electrolyser power consumption at time step $t$ |
| $P_{\text{ElySys},t}$ | Electrolyser system (including compressor) power consumption at time step $t$ |
| $P_{\text{Grid},t}$ | Grid power supply at time step $t$ |

| $P_{\text{nom,Peri}}$ | Nominal power of electrolyser peripherals |
|---|---|
| $P_{\text{Peri},t}$ | Electrolyser peripherals power consumption at time step $t$ |
| $P_{\text{nom,PV}}$ | Nominal power of photovoltaic |
| $P_{\text{PV},t}$ | Photovoltaic power at time step $t$ |
| $P_{\text{RD},t}$ | Redispatch power purchased at time step $t$ |
| $P_{\text{RDava},t}$ | Redispatch power available for purchase at time step $t$ |
| $P_{\text{nom,Stack}}$ | Nominal power of electrolyser stack |
| $P_{\text{Stack},t}$ | Electrolyser stack power consumption at time step $t$ |
| $P_{\text{nom,WindOff}}$ | Nominal power of wind offshore turbine |
| $P_{\text{WindOff},t}$ | Wind offshore turbine power at time step $t$ |
| $P_{\text{nom,WindOn}}$ | Nominal power of wind onshore turbine |
| $P_{\text{WindOn},t}$ | Wind onshore turbine power at time step $t$ |

*Table 2: Optimisation parameters*

| **Parameter** | **Description** |
|---|---|
| $a_{\text{lin},k}$ | Gradient of linearized characteristics curve of electrolyser stack through linearization steps $k$ and $k+1$ |
| $b_{\text{lin},k}$ | The y-axis intersect of linearized characteristics curve of electrolyser stack through linearization steps $k$ and $k+1$ |
| $f_{\text{cap,PV},t}$ | Photovoltaic capacity factor at time step $t$ |
| $f_{\text{cap,WindOff},t}$ | Wind offshore capacity factor at time step $t$ |
| $f_{\text{cap,WindOn},t}$ | Wind onshore capacity factor at time step $t$ |
| $f_{\text{loss,Comp}}$ | Hydrogen loss factor of compressor |
| $\dot{m}_{\text{demand},t}$ | Hydrogen demand mass flow (predefined) at time step $t$ |
| $\Delta t$ | Length of time step in hours |
| $\varepsilon_{\text{Comp}}$ | Specific energy consumption of compressor |
| $\varepsilon_{H_2O}$ | Specific water consumption of electrolyser plant |
| $\varepsilon_{\text{Peri}}$ | Specific energy consumption of electrolyser peripherals |
| $\varepsilon_{\text{nom,Stack}}$ | Specific energy consumption of electrolyser stack at nominal power |
| $\varepsilon_{\text{Stack},k}$ | Specific energy consumption of electrolyser stack at $\frac{k}{K-1} \cdot 100$ (%) of nominal power |

## 7.2  Deep dive result figures calculation

The PPA power share $S_{\text{PPA}}$ in Figure 8 (b) is calculated by dividing the sum of power purchased from all PPAs in the course of the analysed period by the sum of totally purchased power for the operation of the electrolyser.

$$S_{\text{PPA}} = \frac{\sum_{t=1}^{T}(P_{\text{WindOff},t} + P_{\text{WindOn},t} + P_{\text{PV},t})}{\sum_{t=1}^{T} P_{\text{ElySys},t}} \tag{35}$$

The redispatch power share $S_{\text{RD}}$ in Figure 8 (c) is calculated by dividing the sum of purchased redispatch power in the course of the analysed period by the sum of totally purchased power for the operation of the electrolyser.

$$S_{\text{RD}} = \frac{\sum_{t=1}^{T} P_{\text{RD},t}}{\sum_{t=1}^{T} P_{\text{ElySys},t}} \tag{36}$$

The share of redispatch power used $S_{\text{RD,used}}$ in Figure 8 (d) is calculated by dividing the sum of purchased redispatch power in the course of the analysed period by the sum of available redispatch power.

$$S_{\text{RD,used}} = \frac{\sum_{t=1}^{T} P_{\text{RD},t}}{\sum_{t=1}^{T} P_{\text{RDava},t}} \tag{37}$$

The electrolyser utilisation through redispatch usage $U_{\text{Ely,RD}}$ in Figure 8 (e) is calculated by dividing the sum of purchased redispatch power in the course of the analysed period by the nominal size of the electrolyser.

$$U_{\text{Ely,RD}} = \frac{\sum_{t=1}^{T} P_{\text{RD},t}}{P_{\text{nom,Ely}}} \tag{38}$$

The OHSC reductions from redispatch usage $OHSC_{\text{red,RD}}$ in Figure 8 (f) are calculated by dividing the product of the sum of purchased redispatch power and the price of redispatch power purchase by the annual sum of predefined hydrogen demand. Since the redispatch power purchase at positive price levels always causes additional cost, the term is multiplied by minus one to calculate the cost reductions.

$$OHSC_{\text{red,RD}} = \frac{\sum_{t=1}^{T} P_{\text{RD},t} \cdot p_{RD}}{\sum_{t=1}^{T} \dot{m}_{\text{demand},t}} \cdot -1 \tag{39}$$

The OHSC reductions from PPA sizing $OHSC_{\text{red,PPA}}$ in Figure 8 (g) are calculated by dividing the difference between PPA cost in the *PPA* and PPA cost in the *PPA + RD* scenario by the annual sum of predefined hydrogen demand.

$$OHSC_{\text{red,PPA}} =$$
$$\frac{(C_{\text{PP,PV}} + C_{\text{PP,WindOff}} + C_{\text{PP,WindOn}})_{\text{PPA}} - (C_{\text{PP,PV}} + C_{\text{PP,WindOff}} + C_{\text{PP,WindOn}})_{\text{PPA+RD}}}{\sum_{t=1}^{T} \dot{m}_{\text{demand},t} \cdot \Delta t} \tag{40}$$

The OHSC reductions from electrolyser sizing $OHSC_{\text{red,Ely}}$ in Figure 8 (h) are calculated by dividing the difference between annualised electrolyser cost in the *PPA Ref.* scenario and annualised electrolyser cost in the *PPA + RD* scenario by the annual sum of predefined hydrogen demand.

$$OHSC_{\text{red,Ely}} =$$

$$\frac{(C_{\text{TAC,CAPEX,Ely}} + C_{\text{TAC,OPEX,Ely}})_{\text{PPA}} - (C_{\text{TAC,CAPEX,Ely}} + C_{\text{TAC,OPEX,Ely}})_{\text{PPA+RD}}}{\sum_{t=1}^{T} \dot{m}_{\text{demand},t} \cdot \Delta t} \quad (41)$$

The OHSC reductions from storage sizing $OHSC_{\text{red,Ely}}$ in Figure 8 (i) are calculated by dividing the difference between annualised storage cost in the *PPA Ref.* scenario and annualised storage cost in the *PPA + RD* scenario by the annual sum of predefined hydrogen demand.

$$OHSC_{\text{red,Sto}} =$$

$$\frac{(C_{\text{TAC,CAPEX,Sto}} + C_{\text{TAC,OPEX,Sto}})_{\text{PPA}} - (C_{\text{TAC,CAPEX,Sto}} + C_{\text{TAC,OPEX,Sto}})_{\text{PPA+RD}}}{\sum_{t=1}^{T} \dot{m}_{\text{demand},t} \cdot \Delta t} \quad (42)$$

### 7.3 Software
The optimisation problem was implemented in Matlab [40]. Gurobi [41] was used as mathematical solver for all optimisations carried out in this work.

### 7.4 Data and code availability
- All data used, generated or analysed in the course of this study have been deposited at Zenodo [42] and is publicly available at https://doi.org/10.5281/zenodo.15465602. This excludes the raw data provided by EWE Netz GmbH [43] on historical redispatch measures in their distribution network area, due to a non-disclosure agreement.
- The mathematical description of the implemented optimisation problem, including the objective function, all constraints and all parameter assumptions made, is documented in detail and comprehensively in the central article, Methods and Supplementary Information. In addition all original code has been deposited at Zenodo [42] and is publicly available at https://doi.org/10.5281/zenodo.15465602.
- Any additional information required to reanalyse the data reported in this paper is available from the lead contact upon request.

### 7.5 Acknowledgments
This work was funded by the Ministry of Science and Culture of Lower-Saxony (MWK) as part of the innovation laboratory H2-Wegweiser and the German BMBF within the DERIEL project (grant number 03HY122G). The results presented were achieved by computations carried out on the cluster system at the Leibniz Universität Hannover, Germany. We want to thank Michael Claussner for the insightful discussions about "Nutzen statt Abregeln 2.0" and PPA pricing in Germany and Nicolas Stuenkel and Pia Stoelting for their support in collecting and processing historical redispatch data. In addition we want to thank EWE Netz GmbH [43] for the provision of historical data on redispatch measures in their distribution network area.

### 7.6 Author contributions
Brandt, Jonathan (Conceptualization, Software, Analysis, Visualization, Methodology, Writing - original draft, Writing – review & editing)

Bensmann, Astrid; Hanke-Rauschenbach, Richard (Conceptualization, Project administration, Supervision, Writing – review & editing, Funding acquisition)

### 7.7 Corresponding authors
Correspondence to Jonathan Brandt or Astrid Bensmann

## 7.8 Declaration of interests
The authors declare no competing interests.

# C References

Negative redispatch power for green hydrogen production: Game changer or lame duck? A German perspective

# Supplementary material

# Supplementary Note 1: Redispatch time series creation

The used redispatch time series were created based on information about redispatch measures in the German transmission and distribution networks in the years 2022 and 2023. As the provided information about redispatch measures varies between the different network levels, the creation process of time series differs accordingly and is therefore presented separately.

**Transmission level data**

1. The information base on transmission level was obtained from a platform hosted by all German transmission system operators (TSOs), where all redispatch measures of the past years are provided [1].
2. In addition, information on all renewable capacities that are connected on transmission network level and were affected by redispatch measures in the analysed years were obtained from the *Marktstammdatenregister* [2].
3. Then, bias corrected and hourly resolved wind speed data for both years at different heights were obtained for the relevant geographical areas from [3]–[5].
4. The same accounts for time averaged hourly resolved solar radiation data that was obtained for the relevant geographical areas and years from [6].
5. Based on the weather data, hourly resolved historic feed-in time series for all affected photovoltaic (PV), wind onshore and wind offshore capacities were calculated for 2022 and 2023 by using pvlib [7] and windpowerlib [8].
6. Since the available redispatch measure data (see 1.) only provides redispatch information for transformer stations and not for single renewable generation plants, the historic feed-in time series of all respective renewable capacities were cumulated per transformer station based on the connection-information obtained from the *Marktstammdatenregister* (see 2.), resulting in transformer station feed-in time series.
7. Now the transformer station feed-in time series were used to build the redispatch time series based on the redispatch measure data for each transformer station:
    - First: Since the redispatch measure data has a resolution of 15 minutes, the hourly resolved feed-in time series were extended to a 15-minute resolution by linear interpolation.
    - Second: For each documented redispatch measure the designated time slot of the respective feed-in time series was extracted, Supplementary Figure 1 (a) shows an exemplary shortened feed-in time series.
    - Third: The shortened feed-in time series in this specific time slot was cut-off by the maximal power value provided by the redispatch measure data, as shown in Supplementary Figure 1 (b). The resulting top of the cut-off provides a shortened time series of the specific downwards redispatch measure that would be usable in a regional redispatch market. Supplementary Figure 1 (c) shows the time series created in this example.
    - Fourth: Since the used feed-in time series are based on reanalysis weather data, they may differ from the actual occurred feed-in of the respective renewable resources. Therefore, the information on total redispatch energy included in each measure was used to adjust the power cut-off level step-by-step (see: *Energy amount adjustment* in Supplementary Figure 1 (b)), until the energy inside the resulting time series fits the designated energy amount. Provided adjustment was necessary in the first place.

- Fifth: The resulting amount of mini time series (one per redispatch measure) were subsequently accumulated into one redispatch time series per year and transformer station. Thereby, all time steps with no redispatch measure get a zero value.
- Sixth: The 15-minute resolved time series were transformed back to an hourly resolution to fit the regional redispatch market design and the used resolution of the presented optimisation model. Therefore, the average of redispatch power over all four 15-minute periods inside each hour was calculated and assigned to the respective hourly time step.

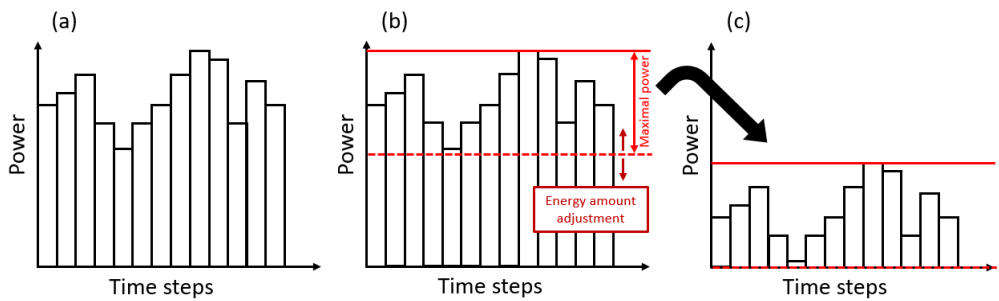

*Supplementary Figure 1: Creation of mini Redispatch time series on transmission network level based on feed-in time series and Redispatch measure information. (a), Feed-in time series in relevant time slot. (b), Cut-off process of feed-in time series based on redispatch measure data and the energy amount adjustment. (c), Shortened time series of the specific downwards redispatch measure usable in a regional redispatch market.*

**Distribution level data**

1. The information base on distribution network level was obtained from platforms hosted by the biggest distribution system operators (DSOs) in northern Germany that cover the majority of network area that is relevant for the analysed relief regions (RR) (German original term: *Entlastungsregionen*) [9]–[11]. The redispatch information on measures of the past years were available from:
    - Avacon Netz GmbH [12]
    - Schleswig-Holstein Netz [13]
    - E.DIS Netz GmbH [14]
    - In addition, EWE Netz GmbH [15] provided already processed hourly resolved redispatch time series for all transformer stations in their network area.
    - WEMAG Netz GmbH [16] and Stromnetz Hamburg GmbH (today: Hamburger Energienetze GmbH) [17] are the only missing DSOs with big network areas relevant for this analysis. Since the available redispatch information provided by WEMAG NETZ GmbH [18] does not allow a clear identification of the affected renewable resources, a creation of respective historic redispatch time series is impossible. As only around 600 and 1300 of all Redispatch measures in 2022 and 2023, in the relevant RR H2 occured in WEMAG NETZ GmbH network area, the created inaccuracy from missing out on WEMAG Netz GmbH data was assessed as acceptably low. As comparison: In the network area of E.DIS Netz GmbH, which is the biggest distribution network operator in H2 approximately 260.000 and 360.000 redispatch measures were reported in 2022 and 2023, respectively. For the network area of Stromnetz Hamburg GmbH no redispatch data was available. Since less than 1 MWh of electricity from renewable sources was produced in Hamburg in both 2022 and 2023, the

missing of this data was assessed as acceptable due to the low relevance for this study [19], [20].

2. As in the process of transmission level data, the steps 2., 3., 4. and 5. were performed to create historic feed-in time series for all renewable capacities involved in a redispatch measure documented in the DSOs data.
3. Now the redispatch measure data for each involved renewable resource was used to modify the associated wind onshore and PV feed-in time series in order to create the redispatch time series:
    o First: The hourly resolved feed-in time series were extended to a 15-minute resolution by linear interpolation.
    o Second: For each documented redispatch measure the designated time slot of the respective feed-in time series was extracted, Supplementary Figure 2 (a) shows an exemplary shortened feed-in time series.
    o Third: The shortened feed-in time series in this specific time slot was cut-off by a percentage of the resource's nominal power provided by the redispatch measure data, as shown in Supplementary Figure 2 (b). The resulting top of the cut-off provides a shortened time series of the specific downwards Redispatch measure that would be usable in a regional Redispatch market. Supplementary Figure 2 (c) shows the time series created in this example. In case the documented measure only covered parts of the 15-minute periods, the magnitude of the measure was adjusted according to the share of affected minutes in the respective 15-minute period.
    o Fourth: The resulting amount of mini time series (one per redispatch measure) were accumulated subsequently into one redispatch time series per year and renewable resource. Thereby, all time steps with no redispatch measure get a zero value.
    o Fifth: The 15-minute resolved time series were transformed back to an hourly resolution to fit the regional redispatch market design and the used resolution of the presented optimisation model. Therefore, the average of redispatch power over all four 15-minute periods inside each hour was calculated and assigned to the respective hourly time step.
4. Since the DSOs redispatch data provides the connection transformer station for each renewable resource, the time series of all renewable resources associated to the same transformer station were merged subsequently to finally get the transformer station redispatch time series on distribution network level for the network areas in 1a-1c (see above).

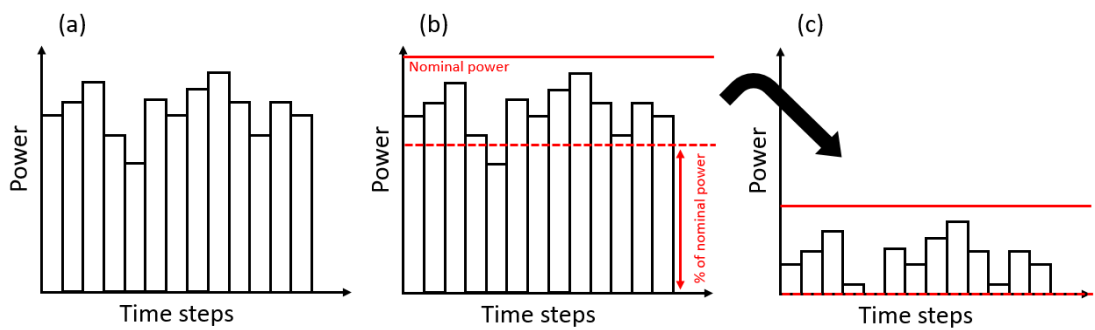

*Supplementary Figure 2: Creation of mini Redispatch time series on distribution network level based on feed-in time series and Redispatch measure information. (a), Feed-in time series in relevant time slot. (b), Cut-off process of feed-in time series based on redispatch measure data. (c), Shortened time series of the specific downwards redispatch measure usable in a regional redispatch market.*

**Regional combination of data**

The generated and provided transformer station redispatch data on transmission and distribution network level were finally merged based on the postal codes of the transformer stations within the respective RRs. The postal codes belonging to each RR were obtained from the published implementation concept of the instrument [21]. The postal codes for all transformer stations were manually searched by using Google Maps [22] based on the names of the transformer stations, which refer to the closest cities.

**Validation of created redispatch data**

Transmission network level:
In order to validate the created redispatch time series, the energy amounts of the time series were compared to the total energy amounts reported within the redispatch measure data.

*Supplementary Table 1: Validation of transmission network level redispatch time series*

| Year | Energy amounts in created time series | Energy amounts reported [1] | Covered percentage |
|---|---|---|---|
| 2022 | 3.769 GWh | 3.811 GWh | 98.897 % |
| 2023 | 6.022 GWh | 6.062 GWh | 99.340 % |

Distribution network level:
Since the redispatch measure data from distribution network operators does not include any information on energy amounts, reports from the German federal network agency were used for validation.

*Supplementary Table 2: Validation of distribution network level redispatch time series*

| Year | Energy amounts in created time series | Total energy amounts reported*[23] | Percentage of redispatch energy on distribution level ** | Calculated Redispatch energy amount on distribution level | Covered percentage |
|---|---|---|---|---|---|
| 2022 | 2.745 GWh | 8.071 GWh | 38 % - 44 % → 41 %*** | 8.071 GWh * 0.41 = 3.309 GWh | 82.956 % |
| 2023 | 3.268 GWh | 10.479 GWh | 38 % - 44 % → 41 %*** | 10.479 GWh * 0.41 = 4.296 GWh | 76.07 % |

*all network levels
**values reported for 2022 and 2023 in individual interim reports by German federal network agency [23]
***Chosen reference value for validation

**Limitations and potential inaccuracies of the time series creation method**

1. As mentioned above the redispatch measure database on distribution network level does not include all relevant network operators. However, due to the aforementioned reasons, the impact on the created redispatch time series and thus on the general conclusions drawn in this study should be marginal.

2. Since windpowerlib [8] does not provide wind turbine models for all relevant wind turbines we used information on hub height and nominal power to find the closest alternative wind turbine model available.
3. An alternative approach to handle the different time resolutions between redispatch data, weather data and market design could have been the disregard of partly affected hours. Since this would have led to an underestimation of redispatch energy amounts, the presented approach including pro rata offsetting of redispatch measures was chosen.
4. As shown in [4], even bias corrected reanalysis weather data can show some inaccuracies when compared with real historic feed-in time series of renewable energy sources. This could lead to inaccuracies in the redispatch time series created for this study. Since the previously presented validation of the created redispatch time series showed high similarities regarding redispatch energy amounts and the chosen method ensured the correct temporal positioning of the measures, the impact of potential inaccuracies was assessed as acceptably low.

# Supplementary Note 2: Technical and economical parameter assumptions

*Supplementary Table 3: Technical and economical parameter assumptions.*

| Component | Parameter | Value | Unit | Reference |
|---|---|---|---|---|
| Wind offshore PPA | Pay-as-produced price* | 0.0883 | €$_{2024}$/kWh | PPA cost-based pricing calculation based on following values. Further details on the calculation method are below this table. |
|  | CAPEX | 3400 | €$_{2024}$/kW | [24] |
|  | OPEX fix | 39 | €$_{2024}$/(kW·a) | [24] |
|  | OPEX var | 0.008 | €$_{2024}$/kWh | [24] |
|  | Lifetime | 25 | a | [24] |
|  | Annual production | 4454 | kWh/kW | Based on used capacity factor time series, see Supplementary Note 3 |
|  | Nominal weighted average cost of capital (WACC) | 8 | % | Oriented on [24] |
| Wind onshore PPA | Pay-as-produced price* | 0.0729 | €$_{2024}$/kWh | Volume weighted average from 2023 and 2024 tenders [25] |
| Photovoltaic PPA | Pay-as-produced price* | 0.0555 | €$_{2024}$/kWh | Volume weighted average from 2023 and 2024 tenders [26] |
| Grid | Electricity price | 0.1976 | €$_{2024}$/kWh | Germany 2024-S1 [27] |
| PEM electrolyser (@30bar output) | CAPEX | 1292.81 | €$_{2024}$/kW | [28] |
|  | OPEX fix | 20.12 | €$_{2024}$/(kW·a) | [28] |
|  | Depreciation time | 15 | a | [28] |
|  | Nominal weighted average cost of capital (WACC) | 9 | % | Calculated from real WACC value in [28], with expected inflation rate of 2 % |
|  | Specific energy consumption at nominal load inclusive peripherals | 52.5 | kWh/kg$H_2$ | [28] |
| | Specific energy consumption peripherals | 3.33 | kWh/kgH$_2$ | [28] |
| | Decrease of specific energy consumption of stack at part load | 1 | % per 10% load reduction | [28] |
| | Specific water consumption | 14 | kgH$_2$O/kgH$_2$ | [28] |
| | Water cost | 3.725 | €$_{2024}$/m$^3$H$_2$O | [28] |
| Compressor (@30bar input & 350bar/30bar*/ 80bar output - pressure tank/salt cavern/pipeline) | CAPEX | 4558.69 | €$_{2024}$/kW | [28] |
| | OPEX fix | 4 | % of CAPEX/a | [28] |
| | Depreciation time | 15 | a | [28] |
| | Nominal weighted average cost of capital (WACC) | 9 | % | Calculated from real WACC value in [28], with expected inflation rate of 2 % |
| | H$_2$ losses | 0.5 | % | [28] |
| | Specific energy consumption for isentropic compression | 1.287/ 0*/ 0.397 | kWh/kgH$_2$ | Own calculation according to [29], [30] |
| | Isentropic efficiency | 80 | % | [31] |
| | Mechanical efficiency | 90 | % | [31] |
| Hydrogen storage – Pressure gas tank (@300 bar) | CAPEX | 730.57 | €$_{2024}$/kg | [28] |
| | OPEX fix | 2 | % of CAPEX/a | [28] |
| | Depreciation time | 25 | a | [28] |
| | Nominal weighted average cost of capital (WACC) | 9 | % | Calculated from real WACC value in [28], with expected inflation rate of 2 % |
| Hydrogen storage – Salt cavern* | CAPEX / Capacity fee* | 12.75** | $\frac{€_{2024}}{kg \cdot a}$ | own calculations** |
| | OPEX var / Usage Fee** | 0.36** | $\frac{€_{2024}}{kg \cdot a}$ | own calculations** |
| | Annuity factor | 1*** | | |
| * PPA prices for onshore wind and photovoltaik are market-based prices oriented on historic tender strike prices [25], [26]. Since tender data for offshore wind do not provide specific strike prices [32], [33], a cost-based pricing approach was chosen to determine the pay-as-produced PPA price. Details on offshore wind PPA price calculation see below. | | | | |

> ** The salt cavern storage is not assumed to be constructed solely for the electrolyser project analysed here but to be a rentable storage bundle. The annual rental fee is thereby composed of a capacity and a usage fee, covering, besides the actual storage of hydrogen, compression and all other above ground measures. Since salt cavern storage capacities are already marketed in this way for natural gas storage, fees from German salt cavern storage operators were respectively modified for hydrogen storage [34].
> *** Assumption made to fit the annual capacity fee into the economical model presented in equation (4) in the Methods in the main document.

**Details on wind offshore PPA price calculation**

To calculate the wind offshore PPA pay-as-produced price, the so called annuity method is used [24]. First the annuity factor $A_{WindOff}$ is calculated as follows, where $r_{in,WindOff}$ is the interest rate and $t_{dep,WindOff}$ the depreciation time. Since we assume long-term PPA contracts with fixed prices rather than indexed prices, the interest rate here displays the nominal weighted average cost of capital.

$$A_{WindOff} = \frac{r_{in,WindOff} \cdot (1 + r_{in,WindOff})^{t_{dep,WindOff}}}{(1 + r_{in,WindOff})^{t_{dep,WindOff}} - 1} \quad (1)$$

Subsequently the PPA price $p_{PPA,WindOff}$ is calculated by adding up the annual capital expenditures $C_{TAC,CAPEX,WindOff}$ and the variable and fixed annual operational expenditures $C_{TAC,OPEXfix,WindOff}$ and $C_{TAC,OPEXvar,WindOff}$ and dividing the sum by the annual production of the wind turbine $P_{WindOff,a}$.

$$p_{PPA,WindOff} = \frac{C_{TAC,CAPEX,WindOff} + C_{TAC,OPEXfix,WindOff} + C_{TAC,OPEXvar,WindOff}}{P_{WindOff,a}} \quad (2)$$

The annual capital expenditures are calculated by multiplying the capital expenditures $C_{CAPEX,WindOff}$ by the annuity factor $A_{WindOff}$.

$$C_{TAC,CAPEX,WindOff} = C_{CAPEX,WindOff} \cdot A_{WindOff} \quad (3)$$

The variable annual operational expenditures are calculated by multiplying the variable operational expenditures $C_{OPEXvar,WindOff}$ by the annual production of the wind turbine $P_{WindOff,a}$.

$$C_{TAC,OPEXvar,WindOff} = C_{OPEXvar,WindOff} \cdot P_{WindOff,a} \quad (4)$$

Finally, the fixed annual operational expenditures are the fixed operational expenditures $C_{OPEXfix,WindOff}$.

$$C_{TAC,OPEXfix,WindOff} = C_{OPEXfix,WindOff} \quad (5)$$

# Supplementary Note 3: Capacity factor time series for PPAs

The capacity factor time series, used in equations (17-19) in the Methods in the main document to map the feed-in of the PPAs, were generated by using bias corrected wind speed data on different heights from [3]–[5] and solar radiation data from [6]. In order to create representative time series for realistic PPAs, the weather data was collected for three different locations in Germany. Once the location of an existing offshore wind park, once the location of an existing onshore wind park and once the location of an existing utility-scale photovoltaic park. Subsequently the weather data was applied to reference plant models using pvlib [7] and windpowerlib [8].

The chosen offshore wind park is DanTysk located in the North sea. Since the wind park is mainly composed of Siemens SWT 120 3.6 wind turbines, the respective model was used as reference plant model to create the offshore wind capacity factor time series.

The chosen onshore wind park is Windpark Niedergoersdorf located in the state of Brandenburg. Since the wind park is mainly composed of Enercon E-115/3000 wind turbines, the respective model was used as reference plant model to create the onshore wind capacity factor time series.

The chosen utility-scale photovoltaic park is Solarpark Templin – Gross Doelln located in the state of Brandenburg. Since the photovoltaic park is south oriented, frameless monocrystalline Solitek modules with a tilt angle of 35 degree and south orientation were used as a reference plant to create the photovoltaic capacity factor time series. In order to balance weather phenomenons of extreme weather years, the capacity factor time series of two representative years where averaged.

# Supplementary Note 4: Additional results *Green hydrogen production exclusively from redispatch power*

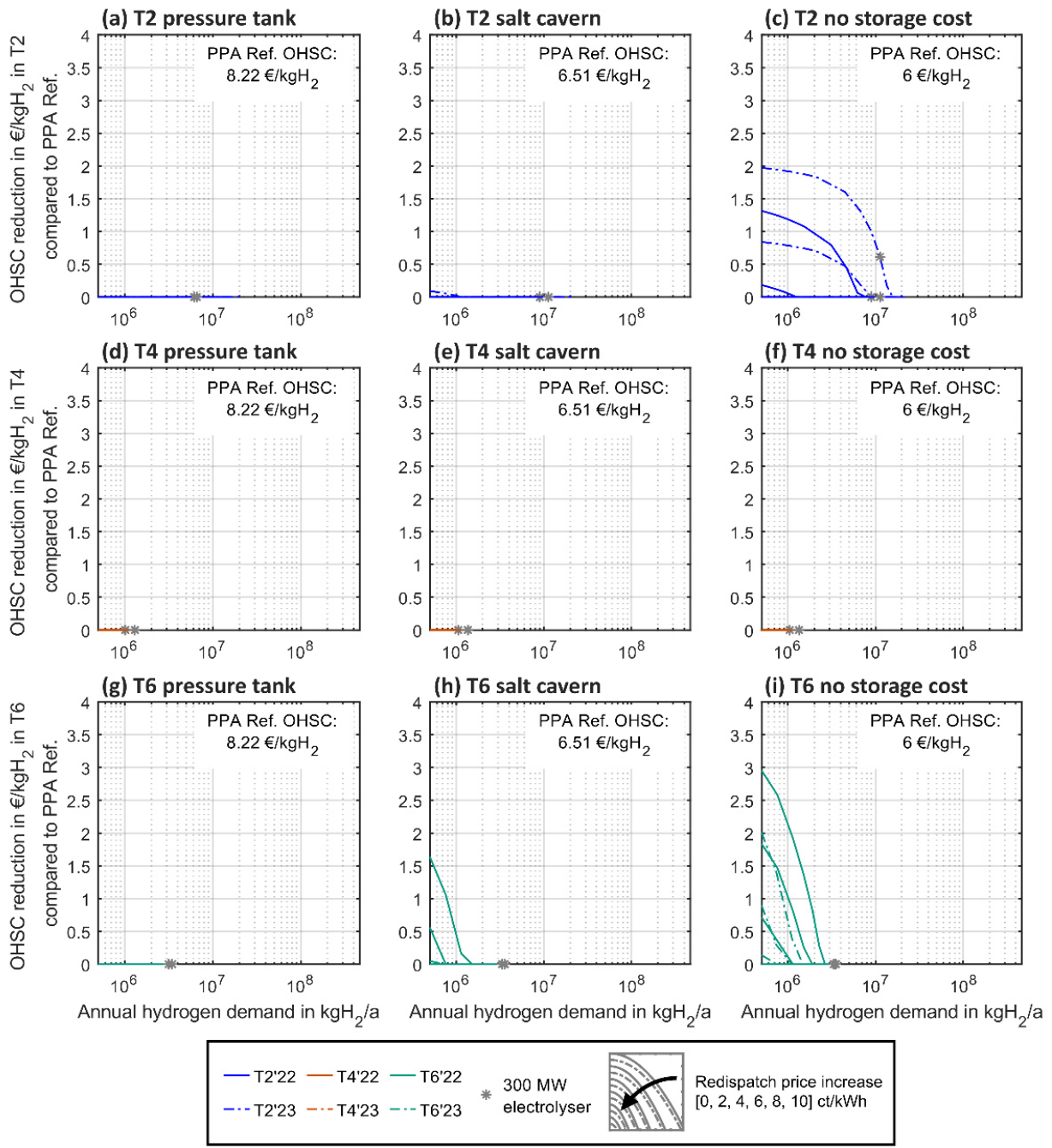

Supplementary Figure 3: Optimisation results for green hydrogen production exclusively from redispatch power for different redispatch price levels and storage options. (a)-(c), OHSC reductions from exclusive redispatch usage compared to exclusive PPA power sourcing in T2. (d)-(f), OHSC reductions from exclusive redispatch usage compared to exclusive PPA power sourcing in T4. (g)-(i), OHSC reductions from exclusive redispatch usage compared to exclusive PPA power sourcing in T6. In addition, the grey star markers show the results for the reference project size of 300 MW. Furthermore, all subfigures show the resulting OHSC of the respective PPA Ref. scenario as a reference.

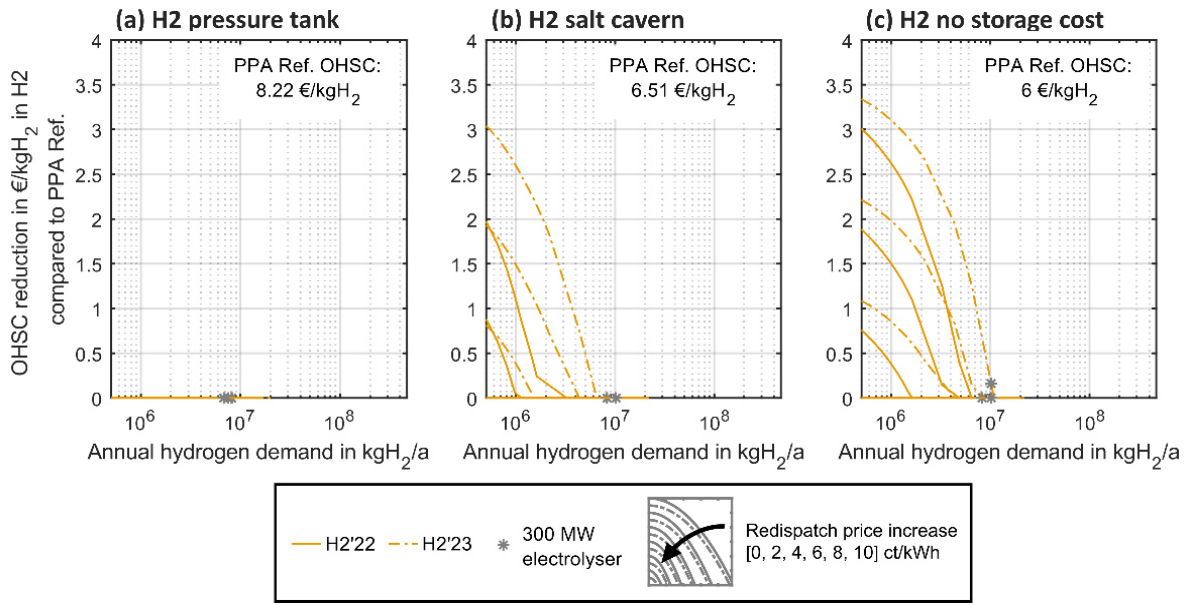

Supplementary Figure 4: Optimisation results for green hydrogen production exclusively from redispatch power for different redispatch price levels and storage options. (a)-(c), OHSC reductions from exclusive redispatch usage compared to exclusive PPA power sourcing in H2. In addition, the grey star markers show the results for the reference project size of 300 MW. Furthermore, all subfigures show the resulting OHSC of the respective PPA Ref. scenario as a reference.

# Supplementary Note 5: Additional results *Combining redispatch and PPA power purchase*

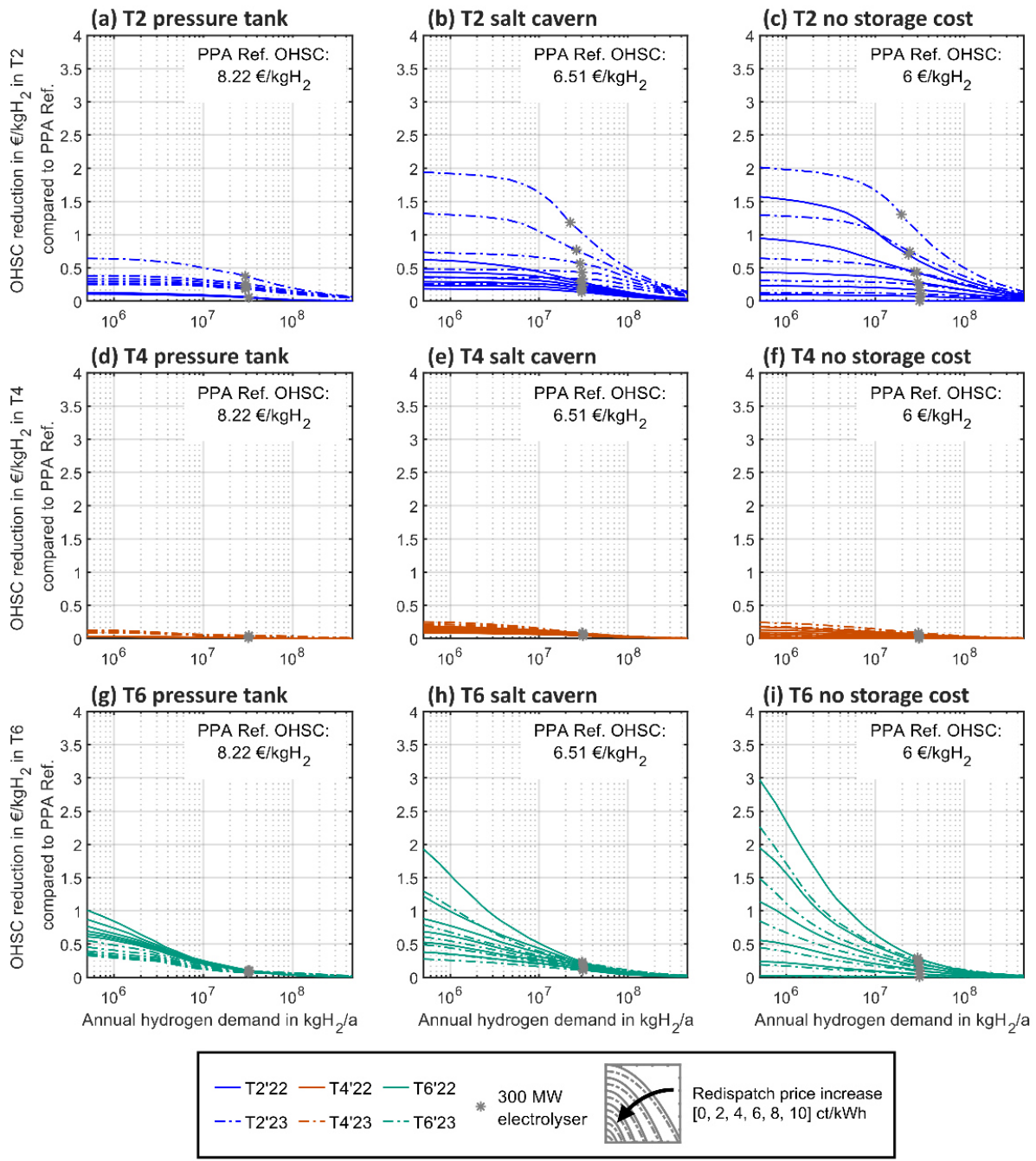

Supplementary Figure 5: Optimisation results for green hydrogen production from combined redispatch and PPA power for different redispatch price levels and storage options. (a)-(c), OHSC reductions from combined PPA and redispatch power usage compared to exclusive PPA power sourcing in T2. (d)-(f), OHSC reductions from combined PPA and redispatch power usage compared to exclusive PPA power sourcing in T4. (g)-(i), OHSC reductions from combined PPA and redispatch power usage compared to exclusive PPA power sourcing in T6. In addition, the grey star markers show the results for the reference project size of 300 MW. Furthermore, all subfigures show the resulting OHSC of the respective PPA Ref. scenario as a reference.

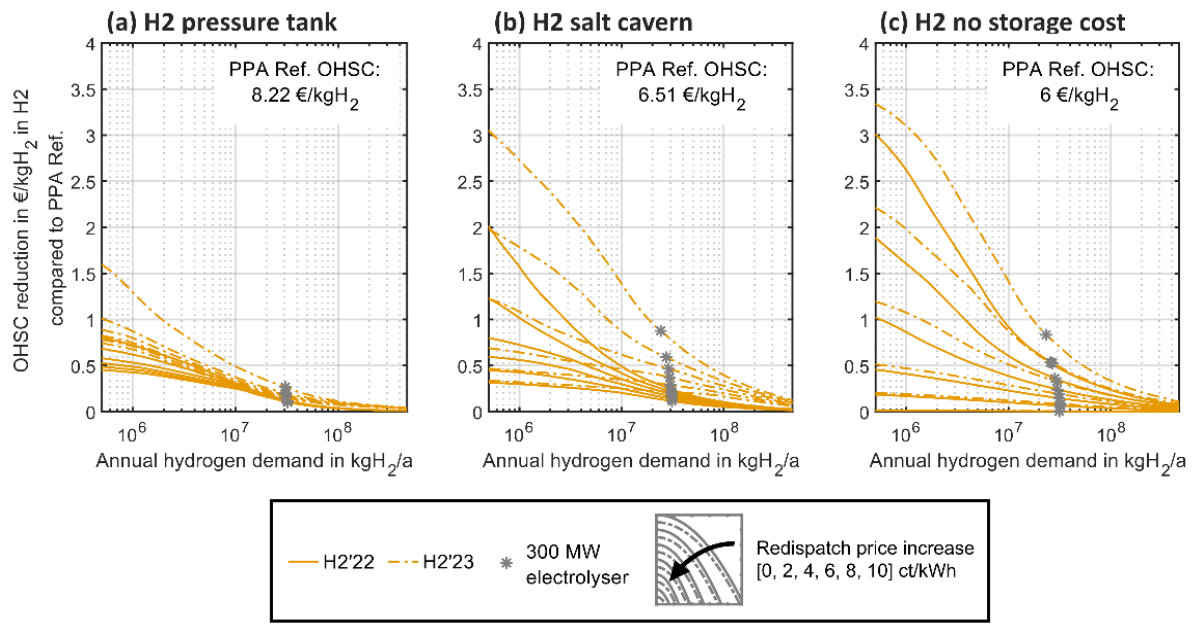

Supplementary Figure 6: Optimisation results for green hydrogen production from combined redispatch and PPA power for different redispatch price levels and storage options. (a)-(c), OHSC reductions from combined PPA and redispatch power usage compared to exclusive PPA power sourcing in H2. In addition, the grey star markers show the results for the reference project size of 300 MW. Furthermore, all subfigures show the resulting OHSC of the respective PPA Ref. scenario as a reference.

# Supplementary Note 6: Additional results *First Mover Projects*

To evaluate whether redispatch power usage is an economically viable option for first mover hydrogen projects, the scenarios *First Mover (FM)* and *FM + RD* are compared. Both cases extend the power purchase options of the scenarios *PPA Ref.* and *PPA + RD* by the option to purchase electricity mix from the grid at the average German electricity price for non-household consumers in 2024 [27]. In the *FM* scenario, the monthly sum of contracted PPA power must equal or exceed the monthly sum of total purchased power for electrolyser operation. In the *FM + RD* scenario, the monthly sum of contracted PPA power must equal or exceed the monthly sum of total purchased power for electrolyser operation minus the purchased redispatch power. In addition to the monthly temporal correlation condition, first mover projects can contract existing PPAs. Their prices are slightly lower than those of newly built PPAs. However, as those prices are difficult to estimate and to ensure a clear assignment of the following results to the monthly temporal correlation condition, the assumed PPA price levels are the same as in all other scenarios. Supplementary Figure 7 depicts the equivalent to Figure 5 and Figure 7 in the main document with OHSC reductions in the *FM + RD* scenario compared to the *FM* scenario.

Comparing the OHSC reductions for low and no storage cost in Supplementary Figure 7 (b), (c), (e), (f), (h), (i) to those in Figure 7 in the main text, only marginal differences from adding electricity mix as additional power purchase option are visible. In contrast, comparing Supplementary Figure 7 (a), (d), and (g) to the equivalent results in Figure 7 in the main text shows a mixed image of possible OHSC reductions for the pressure tank storage option. The differences between the two analysed years in T1 in Supplementary Figure 7 (d) represent this mixed image. At a project size of 300 MW, the maximum OHSC reductions in 2023 (dash-dotted light blue line) are increased from 0.4 €/kgH$_2$ to 0.98 €/kgH$_2$ compared to those without the electricity mix option in Figure 7 in the main text. For the year 2022, on the other hand, the maximum OHSC reductions decrease from 0.84 €/kgH$_2$ to 0.7 €/kgH$_2$. However, despite changes in single high storage cost cases, similar effects and orders of magnitude result for first mover projects compared to the previously analysed scenarios.

Supplementary Figure 8 and Supplementary Figure 9 show the results for all other regions.

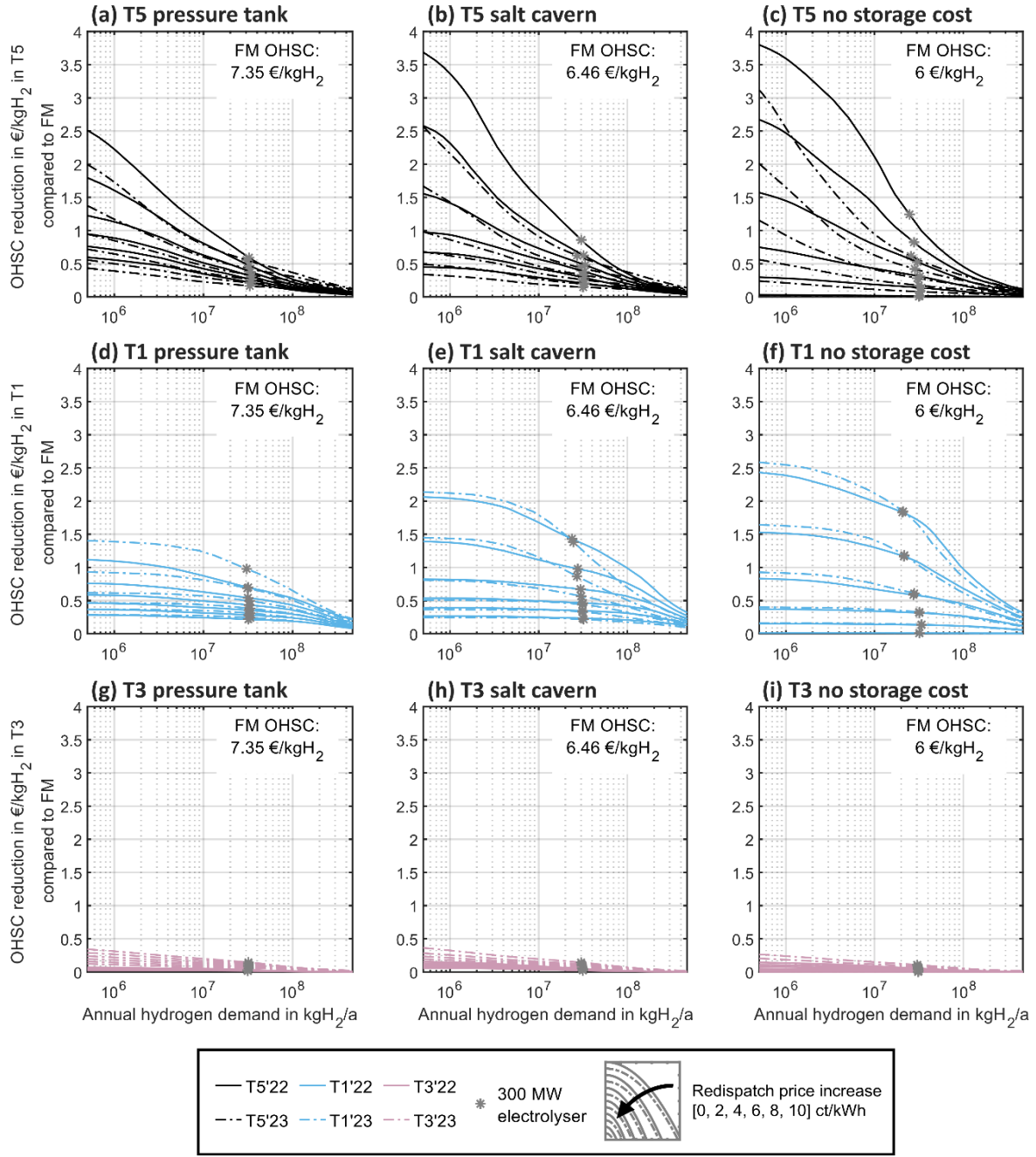

*Supplementary Figure 7: Optimisation results for green hydrogen production of first mover projects from combined redispatch power, PPA power and electricity mix for different redispatch price levels and storage options. (a)-(c), OHSC reductions from combined PPA power, redispatch power and electricity mix usage compared to PPA power and electricity mix sourcing in T5. (d)-(f), OHSC reductions from combined PPA power, redispatch power and electricity mix usage compared to PPA power and electricity mix sourcing in T1. (g)-(i), OHSC reductions from combined PPA power, redispatch power and electricity mix usage compared to PPA power and electricity mix sourcing in T3. In addition, the grey star markers show the results for the reference project size of 300 MW. Furthermore, all subfigures show the resulting OHSC of the respective PPA Ref. scenario as a reference.*

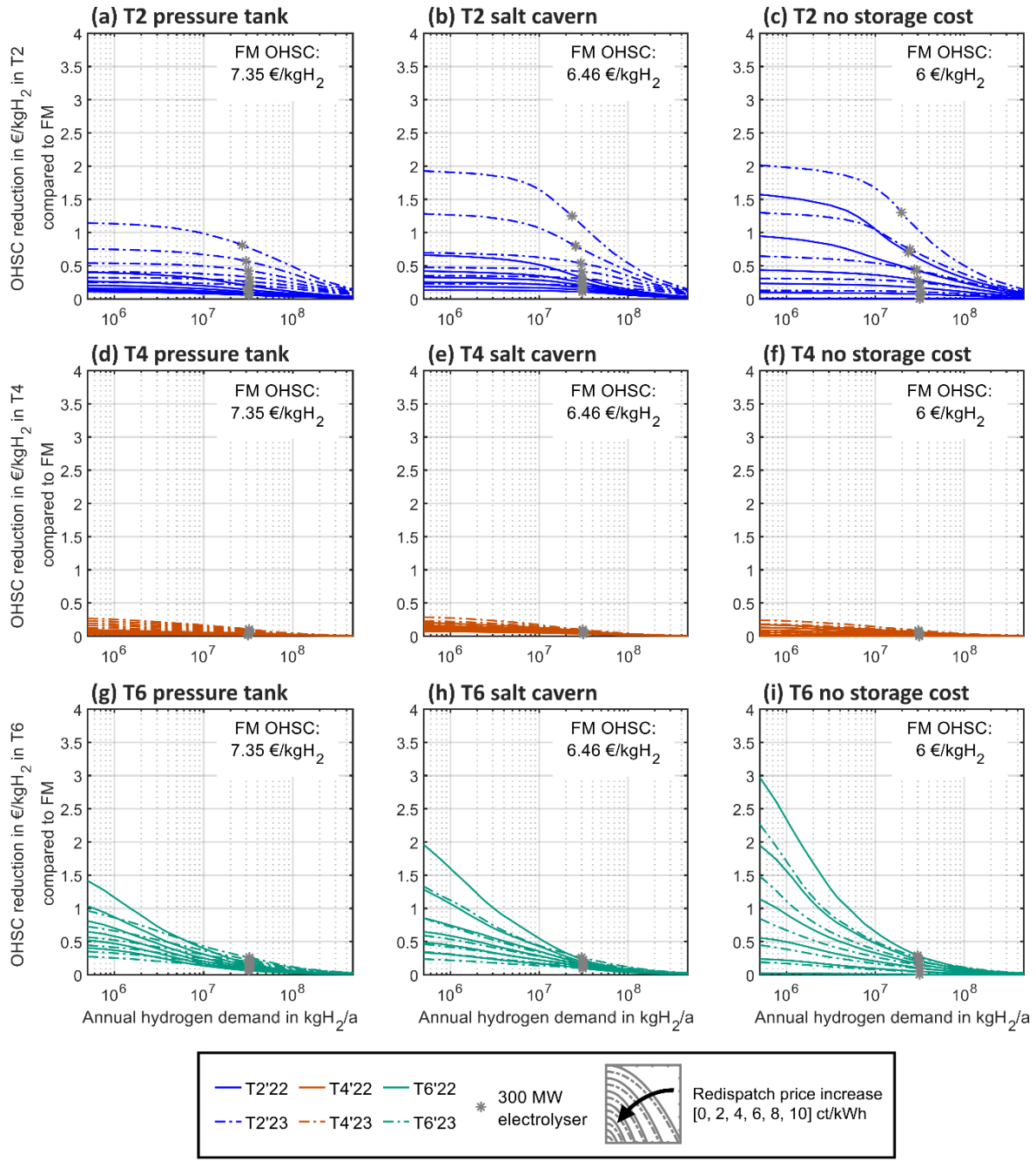

Supplementary Figure 8: Optimisation results for green hydrogen production of first mover projects from combined redispatch power, PPA power and electricity mix for different redispatch price levels and storage options. (a)-(c), OHSC reductions from combined PPA power, redispatch power and electricity mix usage compared to PPA power and electricity mix sourcing in T2. (d)-(f), OHSC reductions from combined PPA power, redispatch power and electricity mix usage compared to PPA power and electricity mix sourcing in T4. (g)-(i), OHSC reductions from combined PPA power, redispatch power and electricity mix usage compared to PPA power and electricity mix sourcing in T6. In addition, the grey star markers show the results for the reference project size of 300 MW. Furthermore, all subfigures show the resulting OHSC of the respective PPA Ref. scenario as a reference.

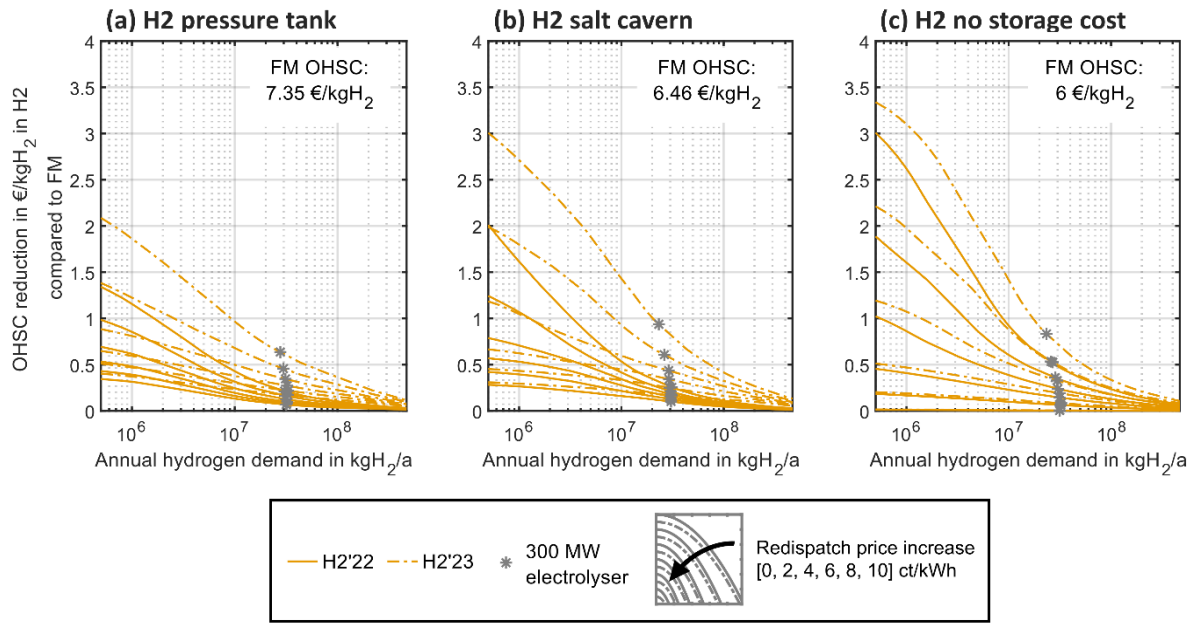

Supplementary Figure 9: Optimisation results for green hydrogen production of first mover projects from combined redispatch power, PPA power and electricity mix for different redispatch price levels and storage options. (a)-(c), OHSC reductions from combined PPA power, redispatch power and electricity mix usage compared to PPA power and electricity mix sourcing in H2. In addition, the grey star markers show the results for the reference project size of 300 MW. Furthermore, all subfigures show the resulting OHSC of the respective PPA Ref. scenario as a reference.